\documentclass[conference,compsoc, anonymous=true]{IEEEtran}
\IEEEoverridecommandlockouts
\IEEEaftertitletext{\vspace{-2\baselineskip}}
\usepackage{hyperref}
\usepackage[dvipsnames]{xcolor}

\hypersetup{
  colorlinks, 
  linkcolor=black,
  urlcolor=RoyalBlue,
  citecolor=black
}

\usepackage{amsmath}
\usepackage{caption}
\usepackage{subcaption}
\usepackage{graphicx}
\usepackage{longtable}
\usepackage{multirow}
\usepackage{booktabs} 
\usepackage{float}
\usepackage{url}
\usepackage{booktabs}
\usepackage{svg}
\graphicspath{{./images/}}
\usepackage{adjustbox}
\usepackage{tabularx}
\usepackage{listings}

\usepackage[table]{xcolor}
\definecolor{RedCell}{HTML}{D32F2F}
\definecolor{OrangeCell}{HTML}{F9A825}
\definecolor{yeargray}{gray}{0.94}

\newcommand{\cellred}{\cellcolor{RedCell}}
\newcommand{\cellorange}{\cellcolor{OrangeCell}}

\definecolor{codegray}{rgb}{0.5,0.5,0.5}
\definecolor{codepurple}{rgb}{0.58,0,0.82}
\definecolor{backcolour}{rgb}{0.95,0.95,0.92}

\usepackage{booktabs,pifont}
\usepackage{xcolor}
\newcommand{\cmark}{\textcolor{green!60!black}{\ding{51}}}
\newcommand{\xmark}{\textcolor{red!70!black}{\ding{55}}}

\usepackage{mdframed}
\def\BibTeX{{\rm B\kern-.05em{\sc i\kern-.025em b}\kern-.08em
    T\kern-.1667em\lower.7ex\hbox{E}\kern-.125emX}}

\usepackage[most]{tcolorbox}
\usepackage{etoolbox} 

\definecolor{takeawaybg}{RGB}{244, 244, 255} 

\newcounter{takeawaycounter}
\tcbset{
  takeaway/.style={
    enhanced,
    colback=takeawaybg,
    colframe=black,
    boxrule=1.5pt,
    arc=4pt,
    left=2pt,
    right=2pt,
    top=2pt,
    bottom=2pt,
    fonttitle=\bfseries,
    before skip=6pt,
    after skip=6pt
  }
}

\newcounter{findingcounter}
\tcbset{
  finding/.style={
    enhanced,
    colback=takeawaybg,
    colframe=black,
    boxrule=1.5pt,
    arc=4pt,
    left=2pt,
    right=2pt,
    top=6pt,
    bottom=2pt,
    fonttitle=\bfseries,
    overlay={
      \node[fill=black, text=white, font=\bfseries, rounded corners=4pt, anchor=west, minimum height=15pt, minimum width=65pt] 
        at ([xshift=10pt,yshift=0pt]frame.north west) {Finding~\thefindingcounter};
    },
    before skip=12pt, after skip=5pt
  }
}

\newcommand{\smartbox}[2]{%
  \ifstrequal{#1}{takeaway}{%
    \refstepcounter{takeawaycounter}%
    \begin{tcolorbox}[takeaway]
      #2
    \end{tcolorbox}
  }{%
    \ifstrequal{#1}{finding}{%
      \refstepcounter{findingcounter}%
      \begin{tcolorbox}[finding]
        #2
      \end{tcolorbox}
    }{%
      \textcolor{red}{Unknown box type!}
    }%
  }%
}

\ifCLASSOPTIONcompsoc
  \usepackage[nocompress]{cite}
\else
  \usepackage{cite}
\fi
\ifCLASSINFOpdf
\else
\fi

\begin{document}
%
\title{Automated Vulnerability Validation and Verification: A Large Language Model Approach}

\author{\IEEEauthorblockN{Alireza Lotfi}
\IEEEauthorblockA{Department of Computer Science\\
Purdue University\\
West Lafayette, IN, USA\\
lotfia@purdue.edu}
\and
\IEEEauthorblockN{Charalampos Katsis}
\IEEEauthorblockA{Department of Computer Science\\
Purdue University\\
West Lafayette, IN, USA\\
ckatsis@purdue.edu}
\and
\IEEEauthorblockN{Elisa Bertino}
\IEEEauthorblockA{Department of Computer Science\\
Purdue University\\
West Lafayette, IN, USA\\
bertino@purdue.edu}}

\maketitle

\begin{abstract}
Software vulnerabilities remain a critical security challenge, providing entry points for attackers to compromise enterprise networks. Despite advances in security practices, the lack of high-quality datasets capturing the behavior of diverse exploits hinders effective vulnerability assessment and mitigation.
This paper introduces an end-to-end multi-step pipeline leveraging generative AI, specifically large language models (LLMs), to address the challenges of orchestrating and reproducing attacks to known software vulnerabilities in controlled environments. Our approach extracts information from CVE disclosures in the National Vulnerability Database, augments it with external public knowledge (e.g., threat advisories, code snippets) using Retrieval-Augmented Generation (RAG), and automates the creation of containerized environments and exploit code tailored to each vulnerability. The pipeline iteratively refines generated artifacts, validates the success of the attack using test cases, and supports complex multi-container setups. Our methodology provides an approach to overcome key obstacles, including noisy and incomplete vulnerability descriptions, by integrating LLMs and RAG to fill information gaps and enhance context. 
We demonstrate the effectiveness of our pipeline across a wide range of vulnerability types, such as memory overflows and denial of service, memory corruption and remote code execution, spanning diverse programming languages, libraries and years. In doing so, we uncover significant inconsistencies in CVE descriptions, emphasizing the need for more rigorous verification of the description in the CVE disclosure process. 
Our approach is model-agnostic, working across multiple LLMs, and we open-source the artifacts to enable reproducibility and accelerate security research. To the best of our knowledge, this is the first system to systematically orchestrate and exploit known vulnerabilities in containerized environments by combining general-purpose LLM reasoning with CVE data and RAG-based context enrichment.
\end{abstract}

%
\IEEEpeerreviewmaketitle





%

\section{Introduction}

Software vulnerabilities have long been a gateway for attackers to infiltrate enterprise networks across various domains. Notable recent examples include the Clop Ransomware Attack, where attackers exploited a zero-day vulnerability in Cleo’s Secure File Transfer products (CVE-2024-50623~\cite{cve_2024_50623}), leading to the extortion of approximately 66 companies following alleged data theft~\cite{checkpoint2024threatreport}, and the Ivanti VPN Exploit (CVE-2025-0282~\cite{cve_2025_0282}), which enabled unauthorized access to corporate networks~\cite{techcrunch2025ivanti}. Despite advances in security practices, software insecurity remains a persistent challenge~\cite{bertino_software}, with thousands of new vulnerabilities disclosed monthly in national vulnerability databases~\cite{NVDdashboard}.

Assessing the impact and potential exploitation of the different software vulnerabilities and how to address them requires a comprehensive understanding of the effects and behaviors of vulnerabilities spanning various types (e.g., resource exhaustion and memory overflows). Different vulnerabilities require different detection and protection strategies~\cite{sharma2023ensemble}. However, a critical barrier to progress in this area is the scarcity and quality of datasets that capture the nuanced behavior of such exploits within software systems~\cite{croft2023data}. Manually orchestrating and replicating attacks to study these behaviors is also a daunting task. It involves creating safe environments for attack execution, setting up appropriate operating systems and libraries, and developing precise exploit code tailored to specific vulnerabilities - tasks that require specialized knowledge. Furthermore, publicly available exploit codes are often unavailable for many vulnerabilities.

\noindent\textbf{Problem Scope.} This work addresses the problem of automated attack reproducibility in isolated environments to advance the understanding of vulnerabilities and enable the design of more effective defensive mechanisms. Specifically, we focus on automating the orchestration of known software vulnerabilities (i.e., CVEs) reported in the National Vulnerability Database. This process entails: (1) generating appropriate containerized environments for controlled attack execution, (2) automating the setup of environments that include operating systems, libraries, vulnerable software executables, and servers, and (3) creating the exploitation code necessary to execute the attacks.

\noindent\textbf{Challenges.} Developing such a system poses several challenges:
\textbf{(C1)} Vulnerabilities are described in natural language, which is often noisy, unstructured and inconsistent, complicating the extraction of key information needed to generate environments and exploits.
\textbf{(C2)} Vulnerability disclosures are frequently poorly written, with minimal contextual details about the attack, making it difficult even for domain experts to fully comprehend the vulnerability, let alone implement the exploit.
\textbf{(C3)} Disclosures typically describe the vulnerability itself but lack details on how the exploit functions or the code required to reproduce it. This creates additional hurdles, such as determining the appropriate programming language for the exploit and identifying the necessary dependencies.

\noindent\textbf{Our Approach.} 
To tackle the aforementioned challenges, we propose a novel pipeline that leverages generative AI, specifically large language models (LLMs). Our multi-step pipeline utilizes off-the-shelf LLMs, such as Chat-GPT 4o, to process vulnerability descriptions expressed in the Common Vulnerabilities and Exposures (CVE) format and automatically generate the execution environments (i.e., Docker containers) and exploitation code necessary to reproduce the reported attacks.

The pipeline begins with a CVE disclosure, structured in JSON format as defined by the CVE Project~\cite{cve_schema}. We employ a carefully designed multi-step prompt that guides the LLM to extract and process key information from the disclosure. This includes identifying device types, product names, affected versions, and other critical details. The prompt also provides instructions on how to handle incomplete or missing information, enabling the model to output a structured representation of the vulnerability for systematic processing in later steps (addressing C1).

Next, we enhance this structured information by employing Retrieval-Augmented Generation (RAG). RAG is an AI technique that enriches the model's contextual knowledge sourced from external resources in addition to the input, significantly improving response accuracy and quality. We use RAG to retrieve additional information from the online resources referenced in the CVE disclosure, such as blogs, advisories, exploit snippets, and CWE assignments. This step achieves two objectives: (1) acquiring context not included in the disclosure and (2) filling gaps, such as missing details about vulnerable functions or parameters. CWEs, in particular, provide valuable insight into the type of vulnerability (e.g., improper input validation) and may include relevant code snippets. This enriched contextual knowledge is then analyzed to establish the pre-conditions (e.g., software versions, dependencies, or operating system configurations) and post-conditions (e.g., segmentation faults, CPU spikes) that comprise the observable state after the attack is executed (addressing C3).

Subsequently, the pipeline aggregates the processed information and generates code into a unified executable script that orchestrates and automates the attack reproduction. We compile each exploit using a language-specific compiler and employ an iterative refinement process that feeds compiler diagnostics back to the model until the code builds cleanly. For C/C++, we combine gdb for precise analysis of build/link issues and runtime faults with AddressSanitizer (ASan)~\cite{address_sanitizer} to detect memory-safety errors missed at compile time. For Python, we rely on ipdb as the counterpart to gdb, providing an interactive debugger.

The pipeline also uses the LLM to determine the number of Docker containers necessary to execute the attack in isolated environments. Depending on the complexity of the CVE, multiple containers may be required. For example, CVE-2025-0665~\cite{cve_2025_0665} requires two containers: one hosting the vulnerable client and another acting as a remote HTTP server. The containers are executed to reproduce the attack, and runtime errors are iteratively addressed with the LLM until the exploit runs successfully. Finally, to verify the success of the attack, we generate test cases based on contextual knowledge. These test cases validate that the observed system state aligns with the vulnerability's expected post-conditions, ensuring accurate exploitation.

\noindent\textbf{Novelty.} To the best of our knowledge, this is the first approach to combine CVE descriptions with context enrichment through RAG to automatically organize, execute, and reproduce attacks in isolated containers, generating detailed attack artifacts. We additionally demonstrate that our pipeline remains operational across multiple publicly available LLMs.

\noindent \textbf{Results.} We evaluated 102 CVEs from 2020–2025 across multiple LLMs and reproduced 71 ($\approx$70\%) CVEs. We covered 55 distinct libraries/projects across 9 programming languages; among the successful cases we generated exploits in 7 languages spanning 40 open-source projects. Our evaluation includes cases with and without public PoCs, showing that our LLM-powered pipeline can reproduce real-world vulnerabilities even when no PoC exists. Across different LLMs, GPT-4o achieved the highest rate of generating correct, complete exploit code.

\noindent\textbf{Ethical Considerations.}
This work focuses exclusively on already-known, publicly disclosed vulnerabilities (CVEs). All experiments and exploit generation were performed in isolated, containerized environments using our own resources, with no interaction with external systems. Our research did not attempt to discover new vulnerabilities or attack live systems.

\noindent\textbf{Contributions.} We make the following contributions:

- Development of an end-to-end pipeline for automated reproduction and orchestration of known software vulnerabilities.

- Integration of CVE disclosures with RAG to aggregate multi-source knowledge orchestration and attack execution in isolated environments.

- Demonstration of our pipeline's effectiveness across a variety of CVEs spanning different CWE types, software systems, and implemented in various programming languages.

- Open-sourcing of our pipeline, generated exploits, and artifacts for all experiments conducted. \url{https://github.com/arlotfi79/CVE_Experiments}

\noindent\textbf{Roadmap.} The paper is structured as follows. Section~\ref{sec:2} provides background information on CVE, their exploitation requirements and the RAG technique. Section~\ref{sec:3} details the steps of our pipeline, while Section~\ref{sec:4} describes the technical aspects of our work, including the LLM-specific challenges encountered and presents an overview of the analyzed CVEs. 
Section~\ref{sec:5} discesses representative challenges encountered during exploit generation, and Section~\ref{sec:6} addresses the limitations of our work. Finally, Section~\ref{sec:7} discusses related work, and Section~\ref{sec:8} concludes the paper, outlining future research directions.  
\section{Background} 
\label{sec:2}
    
\subsection{Common Vulnerabilities and Exposures -- CVE}
The CVE schema standardizes how software and system vulnerabilities are documented and shared, with each record containing key details~\cite{cve_schema}:

\noindent \textbf{CVE ID.} A unique identifier assigned to each vulnerability.
    
\noindent \textbf{Descriptions.} Detailed explanations of the vulnerability, including its nature, impact, and potential exploitation methods.
    
\noindent \textbf{Affected Products.} Information about the software products impacted by the vulnerability, specifying the vendor, product name, and affected versions.
    
\noindent \textbf{Attack Vector.} Details about exploitation requirements, including access type, user interaction, and privilege level.
    
\noindent \textbf{References.} Hyperlinks to external sources, including advisories, reports, or patches related to the vulnerability.
    
\noindent \textbf{Common Weakness Enumeration -- CWE Assignment.} CWE complements CVE by categorizing the root-cause software weaknesses, providing a structured way to analyze recurring patterns and guiding mitigations to prevent repeat vulnerabilities.
    
In parallel, understanding key concepts like pre-conditions and post-conditions is essential for analyzing and exploiting vulnerabilities~\cite{jin2023graphene}. These can be precisely defined as follows: 

\noindent\textbf{Pre-conditions.} Conditions that must be met before exploitation, such as requiring a specific library version or user privilege level in a Linux system.

\noindent\textbf{Post-conditions.} State of the system after the exploit has been executed, such as unauthorized access to sensitive data after successful SQL injection attack.

\subsection{Retrieval Augmented Generation (RAG) for LLMs}
RAG is an advanced AI technique that enhances LLMs by integrating external data sources—both structured and unstructured—that may not have been part of the model’s original training data. This is especially useful for off-the-shelf LLMs, enabling access to proprietary or domain-specific information critical for generating accurate and contextually relevant responses. By bridging pre-trained knowledge with real-time insights, RAG improves accuracy and relevance, particularly in domains requiring up-to-date or specialized information~\cite{nvidia_rag}.

\section{Pipeline overview} \label{sec:3}

\begin{figure*}
  \centering
  \includegraphics[width=.8\textwidth]{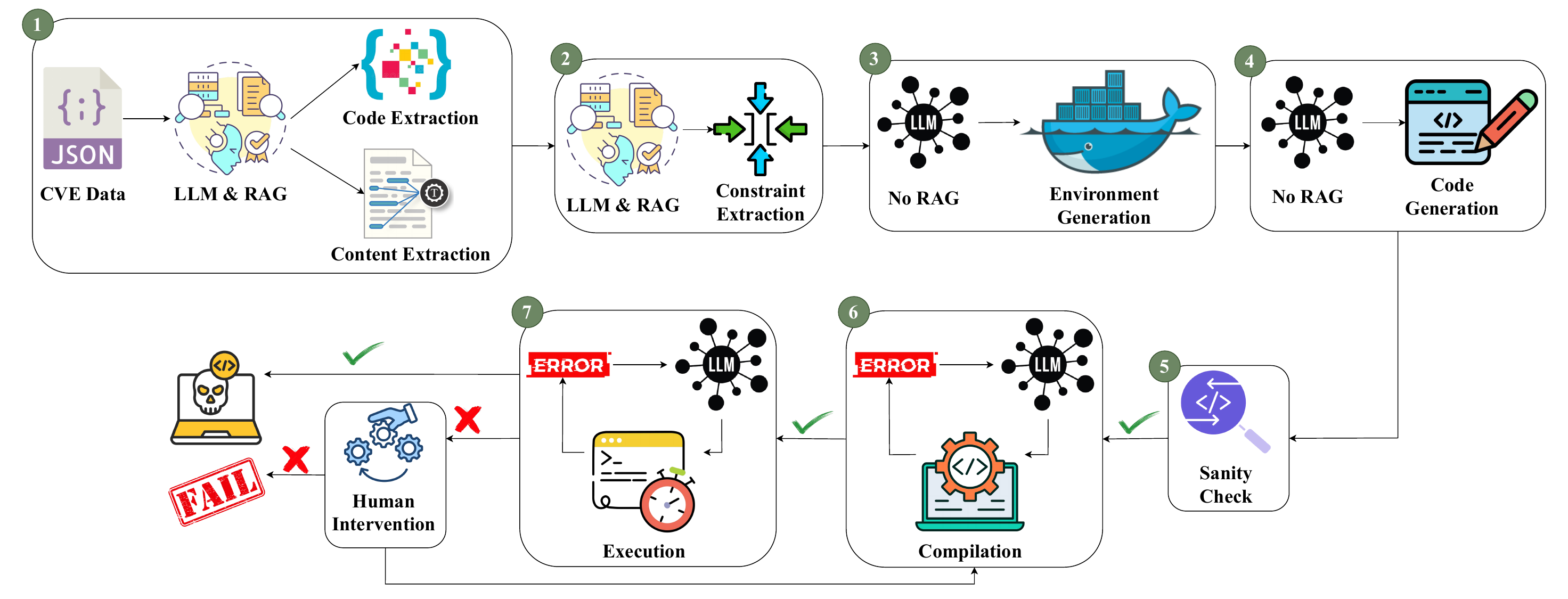}
  \caption{Pipeline Overview} 
  \label{fig:pipeline}
\end{figure*}

This section introduces our pipeline that spans the CVE analysis to exploit code generation, execution, and verification. Figure~\ref{fig:pipeline} shows the structure of the pipeline, highlighting the following key steps:

\noindent \textbf{Pre-Processing.} 
The first step extracts structured information from the CVE using the JSON schema defined by the CVE database~\cite{cve_schema}, including affected products, versions, programming language, and any referenced code snippets. While MITRE’s rules ensure that each CVE contains at least one public reference~\cite{MITRE_CNA_Rules}, our approach does not rely solely on this reference, instead incorporating it when useful but remaining robust in its absence.

To enhance the LLM’s capabilities, we integrate RAG to analyze CWEs referenced in the CVE. This provides a deeper contextual understanding by offering insights into the vulnerability, such as the affected function or library version, and supports the retrieval of code snippets when available. Since CVE~$\rightarrow$~CWE mappings are not always available, our system is designed to function without them: when present, a CWE serves as supplementary context to guide the model in anticipating what the exploit code might look like in case of missing PoC; when absent (e.g., CVE-2021-42576~\cite{cve_2021_42576}), the pipeline relies on descriptions or external sources. The tool is also intended for security professionals, enabling integration of internal threat intelligence to enrich the analysis.

\noindent \textbf{Constraint Extraction.} In this step, we extract the pre-conditions and post-conditions of the vulnerability, which are then used to determine the sequence of actions required for effective exploitation. This sequence subsequently guides the generation of both the execution environment and the corresponding exploit code in later steps.

\noindent \textbf{Environment Generation.} The information extracted earlier is provided to the LLM to advance the process to the next step. The type of product and the vulnerable version are utilized to create an appropriate testing environment in which the exploitation can be executed and verified.

\noindent \textbf{Code Generation.} During this step, our pipeline generates exploit code guided by the configuration and constraints defined earlier, leveraging the extracted pre-conditions, post-conditions, and step-by-step sequence to ensure compatibility with the target environment and reliable triggering of the vulnerability. The resulting configuration and exploit are then consolidated into a single script to streamline testing.

\noindent \textbf{Sanity Check \& Reassessment.} In this step, our pipeline prompts the LLM to perform a thorough review of the generated script to identify and resolve potential issues before testing begins. This helps ensure the vulnerable function is correctly implemented and that all constraints from earlier steps are satisfied.

\noindent \textbf{Compilation.} The exploit is validated by compiling it in the target environment. If errors arise, the sanity check leverages the compiler output to resolve them, and this cycle repeats until successful compilation is achieved.

\noindent \textbf{Execution.} After compilation succeeds, the exploit is executed in the target environment to detect potential runtime issues. Any errors encountered are addressed through an iterative refinement process, and once resolved, the exploit reliably triggers the vulnerability.

\section{Technical details} \label{sec:4}

This section describes the challenges encountered in building our pipeline and details each step from CVE analysis to code generation, with subsections \ref{step:1} through \ref{step:8} highlighting the role of each stage in guiding the LLM toward accurate results. We also present the prompts used to interact with the LLM.

\subsection{Analyzed CVEs} \label{section-CVE-Taxanomy} 
Overall, we examined 55 distinct libraries/projects across 9 programming languages. Among the successful cases, our analysis involves exploits from 7 languages and covers CVEs from 40 unique open-source projects. Table~\ref{tab:cve_vulnerability_taxonomy} summarizes the characteristics of the 71 successfully reproduced CVEs (out of 102 evaluated); the 31 failed cases and brief reasons for failure are reported in Table~\ref{tab:cve_failures_}. Full experimental artifacts for both successful and failed runs can be found in our GitHub repository~\cite{our_github}.

The CVEs were selected according to six criteria: (1) diversity across vulnerability categories to capture different exploit types, (2) coverage of multiple programming languages and execution environments to evaluate generalizability, (3) inclusion of both CVEs with and without publicly available PoCs to assess their impact on exploit generation, (4) a temporal range spanning older, well-studied CVEs as well as very recent disclosures, (5) representation of recent CVEs with limited or poorly written description to test whether contextual gathering through RAG can compensate, and (6) a focus on open-source software and libraries rather than commercial products (e.g., Cisco, Microsoft), given restricted accessibility.
Some CVEs were excluded because they could not be reproduced under our setup, for example when direct hardware access was required and could not be emulated within Docker containers (e.g., CVE-2023-42465~\cite{cve_2023_42465}). 

For each CVE, the table records its unique identifier; the programming language used to implement the exploit; the affected software library or framework; the associated CWE weakness categories; the number of containerized components in the exploit; the number of pipeline iterations executed to obtain a result; whether a PoC exists; whether human intervention was required after the pipeline exhausted its maximum debugging iterations; and whether manual post-condition checks were performed when automated validation was infeasible or unsuccessful (e.g., verifying memory or CPU exhaustion).


\begin{table*}
\centering
\setlength{\tabcolsep}{2pt}
\renewcommand{\arraystretch}{0.95}
\caption{List of successfully reproduced CVEs.}
\label{tab:cve_vulnerability_taxonomy}
\begin{tabular}{r c c p{2.6cm} p{2.2cm} c c c c c}
\toprule
\textbf{No.} & \textbf{CVE ID} & \textbf{Lang} & \textbf{Library} & \textbf{CWE ID} & \textbf{\# Containers} & \textbf{Iter.} & \textbf{PoC} & \textbf{Inter.} & \textbf{Postcond.} \\
\midrule
1  & 2020-1967  & Python     & OpenSSL            & 476                & 2 & 6  & \cmark & \xmark & \xmark \\
\midrule
2  & 2021-29922 & Rust       & net/parser.rs      & 20, 138            & 2 & 4  & \cmark & \xmark & \xmark \\
3  & 2021-31162 & Rust       & Standard Library   & 415                & 1 & 2  & \cmark & \xmark & \xmark \\
4  & 2021-32677 & Python     & FastAPI            & 352                & 3 & 5  & \xmark & \cmark & \xmark \\
5  & 2021-37678 & Python     & TensorFlow         & 502                & 1 & 1  & \xmark & \xmark & \xmark \\
6  & 2021-38191 & Rust       & Tokio              & 362                & 1 & 3  & \cmark & \xmark & \xmark \\
7  & 2021-42576 & Python     & bluemonday         & --                 & 1 & 4  & \cmark & \xmark & \xmark \\
\midrule
8  & 2022-0391  & Python     & urllib             & 74                 & 1 & 4  & \cmark & \xmark & \xmark \\
9  & 2022-3358  & C          & OpenSSL            & 476                & 1 & 6  & \xmark & \xmark & \xmark \\
10 & 2022-22817 & Python     & Pillow             & --                 & 1 & 6  & \xmark & \xmark & \xmark \\
11 & 2022-25219 & C          & OpenSSL            & --                 & 1 & 8  & \xmark & \xmark & \xmark \\
12 & 2022-25235 & C          & Expat              & 116                & 1 & 2  & \xmark & \xmark & \xmark \\
13 & 2022-25315 & C          & Expat              & 190                & 1 & 10 & \cmark & \cmark & \xmark \\
14 & 2022-35934 & Python     & TensorFlow         & 617                & 1 & 1  & \cmark & \xmark & \xmark \\
15 & 2022-35986 & Python     & TensorFlow         & 20                 & 1 & 2  & \cmark & \xmark & \xmark \\
16 & 2022-41883 & Python     & TensorFlow         & 125                & 1 & 1  & \cmark & \xmark & \xmark \\
17 & 2022-41885 & Python     & TensorFlow         & 131                & 1 & 3  & \cmark & \xmark & \xmark \\
18 & 2022-48560 & Python     & heapq              & 416                & 1 & 2  & \cmark & \xmark & \cmark \\
\midrule
19 & 2023-0217  & C          & OpenSSL            & 476                & 1 & 5  & \xmark & \xmark & \xmark \\
20 & 2023-5678  & C          & OpenSSL            & 606, 754           & 1 & 4  & \xmark & \xmark & \xmark \\
21 & 2023-22809 & Shell      & Sudo               & 269                & 1 & 6  & \cmark & \cmark & \xmark \\
22 & 2023-25664 & Python     & TensorFlow         & 120, 122           & 1 & 1  & \cmark & \xmark & \xmark \\
23 & 2023-25665 & Python     & TensorFlow         & 476                & 1 & 3  & \cmark & \xmark & \xmark \\
24 & 2023-25666 & Python     & TensorFlow         & 697                & 1 & 8  & \cmark & \xmark & \xmark \\
25 & 2023-25667 & Python     & TensorFlow         & 190                & 2 & 3  & \cmark & \xmark & \xmark \\
26 & 2023-25668 & Python     & TensorFlow         & 122, 125           & 1 & 3  & \cmark & \xmark & \xmark \\
27 & 2023-25669 & Python     & TensorFlow         & 697                & 1 & 1  & \cmark & \xmark & \xmark \\
28 & 2023-25672 & Python     & TensorFlow         & 476                & 1 & 4  & \cmark & \xmark & \xmark \\
29 & 2023-25674 & Python     & TensorFlow         & 476                & 1 & 4  & \cmark & \xmark & \xmark \\
30 & 2023-25675 & Python     & TensorFlow         & 697                & 1 & 2  & \cmark & \xmark & \xmark \\
31 & 2023-25801 & Python     & TensorFlow         & 415                & 1 & 3  & \cmark & \xmark & \xmark \\
32 & 2023-28487 & Shell      & Sudo               & 116                & 1 & 5  & \xmark & \cmark & \xmark \\
33 & 2023-33966 & JavaScript & Deno               & 269, 276           & 2 & 2  & \xmark & \xmark & \xmark \\
34 & 2023-37658 & Python     & fast-poster        & 79                 & 2 & 2  & \cmark & \xmark & \xmark \\
35 & 2023-39325 & Go         & HTTP/2 Server      & 400                & 2 & 5  & \xmark & \xmark & \xmark \\
36 & 2023-49210 & JavaScript & OpenSSL            & 77                 & 1 & 2  & \cmark & \xmark & \xmark \\
37 & 2023-49292 & Go         & Crypto             & 200                & 1 & 6  & \xmark & \xmark & \xmark \\
38 & 2023-6507  & Python     & subprocess         & 269                & 1 & 3  & \cmark & \xmark & \xmark \\
\midrule
39 & 2024-0397  & Python     & SSL                & 362                & 1 & 6  & \xmark & \xmark & \xmark \\
40 & 2024-12224 & Rust       & rust-url           & 352, 1289          & 1 & 4  & \cmark & \cmark & \xmark \\
41 & 2024-1597  & Java       & PostgreSQL         & 89                 & 2 & 10 & \cmark & \cmark & \xmark \\
42 & 2024-24789 & Go         & archive/zip        & 390                & 2 & 10 & \xmark & \cmark & \xmark \\
43 & 2024-24790 & Go         & netip              & 180                & 1 & 3  & \xmark & \xmark & \xmark \\
44 & 2024-27304 & Go         & pgx                & 89, 190            & 3 & 10 & \xmark & \xmark & \xmark \\
45 & 2024-27934 & JavaScript & Deno               & 416                & 1 & 7  & \cmark & \xmark & \xmark \\
46 & 2024-27936 & JavaScript & Deno               & 150                & 1 & 3  & \cmark & \xmark & \xmark \\
47 & 2024-33664 & Python     & Crypto             & 400                & 1 & 11 & \cmark & \xmark & \xmark \\
48 & 2024-4032  & Python     & ipaddress          & 697                & 1 & 6  & \cmark & \xmark & \xmark \\
49 & 2024-40627 & Python     & FastAPI            & 204                & 2 & 6  & \cmark & \xmark & \xmark \\
50 & 2024-43204 & Python     & Apache Server      & 918                & 3 & 5  & \xmark & \xmark & \xmark \\
51 & 2024-4340  & Python     & sqlparse           & 674                & 1 & 1  & \cmark & \xmark & \xmark \\
52 & 2024-5991  & C          & wolfSSL            & 125                & 1 & 10 & \xmark & \xmark & \cmark \\
53 & 2024-9287  & Python     & venv               & 77, 428            & 1 & 2  & \cmark & \xmark & \xmark \\
\midrule
54 & 2025-0182  & Python     & danswer-ai         & 400                & 2 & 3  & \xmark & \xmark & \cmark \\
55 & 2025-0665  & C          & curl               & 1341               & 2 & 4  & \cmark & \xmark & \xmark \\
56 & 2025-21620 & JavaScript & Deno               & 200                & 3 & 3  & \cmark & \xmark & \xmark \\
57 & 2025-22153 & Python     & RestrictedPython   & 843                & 1 & 2  & \xmark & \xmark & \xmark \\
58 & 2025-22874 & Go         & crypto/x509        & --                 & 2 & 4  & \xmark & \xmark & \xmark \\
59 & 2025-24015 & JavaScript & Deno               & 347                & 1 & 5  & \cmark & \xmark & \xmark \\
60 & 2025-24898 & Rust       & OpenSSL            & 416                & 2 & 7  & \cmark & \xmark & \xmark \\
61 & 2025-27498 & Rust       & aes-gcm            & 347                & 1 & 2  & \cmark & \xmark & \xmark \\
62 & 2025-29744 & JavaScript & pg-promise         & 89                 & 2 & 8  & \cmark & \xmark & \xmark \\
63 & 2025-30223 & Go         & Beego              & 79                 & 1 & 4  & \cmark & \xmark & \xmark \\
64 & 2025-32462 & Shell      & Sudo               & 863                & 2 & 6  & \cmark & \xmark & \xmark \\
65 & 2025-4516  & Python     & CPython            & 416                & 1 & 1  & \xmark & \xmark & \cmark \\
66 & 2025-47268 & C          & iputils            & 190                & 2 & 5  & \cmark & \xmark & \xmark \\
67 & 2025-48054 & JavaScript & radish             & 1321               & 1 & 2  & \xmark & \xmark & \xmark \\
68 & 2025-48754 & Rust       & memory\_pages      & 369                & 1 & 3  & \cmark & \xmark & \xmark \\
69 & 2025-48755 & Rust       & sdk                & 762                & 1 & 2  & \cmark & \xmark & \xmark \\
70 & 2025-48888 & JavaScript & Deno               & 863                & 1 & 2  & \cmark & \xmark & \xmark \\
71 & 2025-48934 & JavaScript & Deno               & 201                & 1 & 1  & \cmark & \xmark & \xmark \\
\bottomrule
\end{tabular}
\end{table*}

\begin{table*}
\centering
\setlength{\tabcolsep}{3pt}
\renewcommand{\arraystretch}{0.95}
\caption{List of Failed CVEs}
\label{tab:cve_failures_}
\begin{tabular}{r c c c p{8.8cm}}
\toprule
\textbf{No.} & \textbf{CVE ID} & \textbf{Language} & \textbf{Library} & \textbf{Reason of Failure} \\
\midrule
1  & 2021-4034   & C          & Polkit/GLib                & Non-Verifiable CVEs \\
2  & 2021-28032  & Rust       & Nano\_arena                & Environment Setup Issues \\
3  & 2021-28875  & Rust       & Standard Library           & Code Generation Issues \\
4  & 2021-28878  & Rust       & Standard Library           & Code Generation Issues \\
5  & 2021-44228  & Java       & Apache Log4j               & Environment Setup Issues \\
6  & 2022-2274   & C          & OpenSSL                    & Non-Verifiable CVEs \\
7  & 2022-22816  & Python     & Pillow                     & Code Generation Issues \\
8  & 2022-24713  & Rust       & Regex                      & Code Generation Issues \\
9  & 2022-35938  & Python     & TensorFlow Lite Micro      & Environment Setup Issues \\
10 & 2022-35939  & Python     & TensorFlow Lite            & Environment Setup \& Code Generation Issues \\
11 & 2022-36087  & Python     & OAuthLib                   & Code Generation Issues \\
12 & 2022-47069  & Python     & p7zip                      & Code Generation Issues \\
13 & 2023-25663  & Python     & TensorFlow                 & Non-Verifiable CVEs \\
14 & 2023-25670  & Python     & TensorFlow                 & Environment Setup Issues \\
15 & 2023-28446  & JavaScript & Deno                       & Code Generation Issues \\
16 & 2023-42465  & Shell      & Sudo                       & Non-Verifiable CVEs \\
17 & 2023-45827  & JavaScript & Dot Diver                  & Code Generation Issues \\
18 & 2023-46133  & JavaScript & CryptoES                   & Environment Setup Issues \\
19 & 2024-2410   & C++        & Protobuf                   & Code Generation Issues \\
20 & 2024-6874   & C          & Libcurl (macidn backend)   & Non-Verifiable CVEs \\
21 & 2024-27931  & JavaScript & Deno                       & Environment Setup \& Code Generation Issues \\
22 & 2024-32468  & JavaScript & Deno                       & Code Generation Issues \\
23 & 2024-44905  & Go         & Go-pg                      & Code Generation Issues \\
24 & 2024-44906  & Go         & Pgdriver                   & Code Generation Issues \\
25 & 2024-53319  & C++        & Qualisys SDK               & Code Generation Issues \\
26 & 2025-4574   & Rust       & Crossbeam-channel          & Environment Setup \& Code Generation Issues \\
27 & 2025-5791   & Rust       & Crate                      & Code Generation Issues \\
28 & 2025-32433  & Erlang     & Erlang/OTP                 & Environment Setup \& Code Generation Issues \\
29 & 2025-43855  & JavaScript & tRPC                       & Environment Setup \& Code Generation Issues \\
30 & 2025-47273  & Python     & Setuptools                 & Code Generation Issues \\
31 & 2025-53605  & Rust       & Protobuf                   & Environment Setup \& Code Generation Issues \\
\bottomrule
\end{tabular}
\end{table*}

\subsection{Challenges} \label{subsec:4-1}
Designing our pipeline required addressing the following challenges:

\noindent \textbf{Censorship.} Publicly available LLMs are censored, refusing to generate code exploits for CVE disclosures~\cite{VER-MALLA,llm-cencorship}.  We can bypass this limitation of LLMs using jailbreak techniques, which we discuss in Step~\ref{step:4}.

\noindent \textbf{Memory-constrained LLMs.} LLMs tend to forget or become confused when given too much data~\cite{llm-foregetting,openai_forgetting}. In each step of our pipeline, the prompts consistently include repeated information, such as the CVE description and outputs from previous steps. This repetition ensures that the LLM maintains context and has the necessary data to generate accurate outputs for each specific step.

\noindent \textbf{LLM Context Size.} LLMs have a fixed context window per turn (prompt + retrieved text + system prompt). With GPT-4o (OpenAI Tier 3) we benefited from a larger context window and higher tokens-per-minute limit (800K TPM), but still cap outputs at 4,096 tokens, aligning each pipeline step with this size to maintain efficiency. Early overlong prompts were fixed by splitting the prompt into Steps~\ref{step:1}–\ref{step:8}, keeping prompts manageable without carrying the full history. Debugging was the only step with substantial log size (peaks $\approx$480 MB); to control size, regex filters forward only essentials—errors, warnings, Docker build/run failures, and pertinent code snippets.

\noindent \textbf{LLM Hallucination.} Hallucination occurs when an LLM produces fluent yet inaccurate outputs. Adam et al.~\cite{llm_hallucination} attribute this to model's training and evaluation processes that favor confident guessing over expressing uncertainty. In our pipeline, hallucination is rare during early steps (extracting constraints, vulnerable library, code, and environment) and appears mainly in debugging, where the model sees nearly the same code across iterations; lacking a clear fix, it may drift into irrelevant outputs. We mitigate this by constraining debugging with explicit pre/post-conditions and a system prompt listing past mistakes (defined in the debugging prompt). Additionally, verification (Step~\ref{step:5}) catches errors early, preventing invalid inputs from reaching debugging and reducing re-runs. To mitigate misdirected starts, a three-attempt limit per CVE explores distinct strategies, increasing the chance of fewer iterations to a working exploit or success where earlier attempts failed.

\noindent \textbf{Complexity.} Code generation is a non-trivial task in complex scenarios, often requiring simultaneous consideration of multiple dependencies. In such cases, the model is prompted to identify the number of containers required for successful exploitation. Details of how this process works are given in Step \ref{step:3}.

\noindent \textbf{Code Compilability.} The generated code is not always compilable, leading to errors and warnings~\cite{CG-Refining}. To ensure that the generated code is compilable, three separate steps are implemented: \ref{step:5}, \ref{step:6-7}, and \ref{step:8}. Step~\ref{step:5} verifies the correctness of the code with the LLM without compiling it. Step~\ref{step:6-7} compiles the code and checks for errors or warnings. Finally, step~\ref{step:8} uses an iterative approach with the LLM to address and resolve any issues.

\noindent \textbf{Expected Functionality.} LLM-generated code does not always meet the intended behavior: it may compile cleanly yet fail to reach the vulnerable path or satisfy the pre or post-conditions to trigger the vulnerability. To reduce such false positives, Step~\ref{step:4} requires the model to produce concrete test cases and explicit success checks. Even so, the model can still ``simulate'' success. In those cases, a human might need to review the code and test oracles, or restart the run as needed to resolve the issue.

\subsection{Step 1 - Pre-Processing} \label{step:1}
\noindent\textbf{(1) Content Extraction.} As noted in Subsection~\ref{subsec:4-1}, context-window limits constrain how much text an LLM can process per turn, especially as prompts grow across later steps. To address this, we extract only the essential information for exploitation from the CVE JSON file, reducing input size and avoiding overflows. Listing~\ref{steps:1} shows the extraction prompt, and Listing~\ref{fig:pipeline_example-step1-1} illustrates the output for CVE-2021-29922~\cite{cve_2021_29922}.
\begin{lstlisting}[caption={CVE-2021-29922 Step 1.1}, label={fig:pipeline_example-step1-1}]
{"Category": "Application",
  "Subcategory": "Library",
  "Product": {"Name": "Rust", "Vendor": "rust-lang", "Version": ["<1.53.0"]}
}
\end{lstlisting}

\noindent\textbf{(2) Language \& Code Extraction.} Some CVEs lack sufficient detail for direct exploitation. In such cases, we retrieve contextual information from reliable sources, particularly the references embedded in CVE disclosures, which often include advisories, code snippets, or PoCs. For instance, CVE-2023-25668~\cite{cve_2023_25668} is poorly documented, yet its references include a PoC that clarifies how the flaw can be exploited.

This retrieval process occurs in two sub-steps: first, the relevant programming language for exploitation is identified; second, all references are examined for vulnerable code listings that could guide PoC generation. Rather than prioritizing individual sources, our pipeline automatically gathers this information through the model's built-in RAG feature, which also searches beyond CVE references (e.g., Internet, GitHub, or other online repositories). When references are scarce, reproduction becomes challenging for both humans and models; nevertheless, the pipeline can still attempt exploitation with limited data. For example, our pipeline successfully reproduces CVE-2024-5991~\cite{cve_2024_5991} despite the absence of any PoC. Listing~\ref{steps:1-2} shows how the model is prompted in this step, while Listing~\ref{fig:pipeline_example-step1-2-3} illustrates a sample output for CVE-2021-29922~\cite{cve_2021_29922}.

\begin{lstlisting}[caption={CVE-2021-29922 Step 1.2}, label={fig:pipeline_example-step1-2-3}]
{"Language": "Rust",
  "Vulnerable Code Snippets": [
    {"Code": "use std::net::IpAddr;\nuse std::str::FromStr;\n\nfn main() {\n    let ip = IpAddr::from_str(\"010.8.8.8\").unwrap();\n    println!(\"Parsed IP: {}\", ip);\n}",
  "Link": "https://github.com/sickcodes/security/blob/master/advisories/
  SICK-2021-015.md"}]
}
\end{lstlisting}

\subsection{Step 2 - Constraint Extraction}
Setting up an accurate exploitation environment requires first identifying all prerequisites and then verifying the exploit’s effectiveness by examining the resulting system or program state. This verification clarifies the pre-conditions, post-conditions, and sequence of actions needed to trigger the vulnerability.

To support this process, we prompt the model with the information extracted in Step~\ref{step:1}, enabling a deeper understanding of the vulnerability. In addition, the model is instructed to retrieve contextual details from the CWE MITRE website, as outlined in line 3 of Listing~\ref{steps:2}. Because CWE entries share a consistent URL structure, differing only by CWE number. This step supplies supplementary context that improves the accuracy of the generated exploitation code. The corresponding prompt is shown in Listing~\ref{steps:2}, and Listing~\ref{fig:pipeline_example-step2} presents a sample output for CVE-2023-25674~\cite{cve_2023_25674}.

\begin{lstlisting}[caption={CVE-2021-29922 Step 2}, label={fig:pipeline_example-step2}]
{"Vulnerability Title": "Null Pointer Dereference in TensorFlow's RandomShuffle with XLA Enabled",
  "CWEs Description": [
    {"CWE-ID": "CWE-476",
     "Description": "The product dereferences a pointer that it expects to be valid but is NULL."}],
  "Preconditions": [
    "TensorFlow version prior to 2.12.0 or 2.11.1 is installed.",
    "XLA (Accelerated Linear Algebra) compilation is enabled.",
    "The `RandomShuffle` operation is invoked with specific parameters that lead to a null pointer dereference."],
  "Postconditions": [
    "The application crashes due to a null pointer dereference.",
    "Potential denial of service as the process terminates unexpectedly."],
  "Language": {"Language": "Python"},
  "Steps": [
    "Import TensorFlow: `import tensorflow as tf`",
    "Define the `RandomShuffle` operation: `func = tf.raw_ops.RandomShuffle`",
    "Set parameters that trigger the vulnerability: `para = {'value': 1e+20, 'seed': -4294967297, 'seed2': -2147483649}`",
    "Define a TensorFlow function with JIT compilation enabled: `@tf.function(jit_compile=True) def test(): y = func(**para); return y`",
    "Invoke the function: `test()`"]}
\end{lstlisting}

\subsection{Step 3 - Environment Generation} \label{step:3}
A reliable testing environment is essential for validating any exploit. Our pipeline employs an automated, LLM-guided process that generates a tailored containerized Docker setup. Docker provides key benefits: it is infrastructure-independent, portable, and ensures consistent, isolated environments with only the required libraries. This guarantees reproducibility and simplifies artifact verification, avoiding the limitations of approaches that merely list dependencies without offering an executable setup.

\noindent \textbf{(1) Required Number of Docker Containers.} Some CVEs cannot be reproduced in a single container because they depend on interactions between multiple services. For example, CVE-2025-0665 requires two containers: one hosting the vulnerable client and another acting as a remote HTTP server. Determining the correct number of containers is therefore a critical step. In our experiments, 20 of 71 successfully reproduced CVEs ($\approx28\%$) required multi-container setups. The corresponding prompt is shown in Listing~\ref{steps:4-1}, and Listing~\ref{fig:pipeline_example-step3-1} shows a sample output produced by our prompt for this CVE.

\begin{lstlisting}[caption={CVE-2021-29922 Step 3.1}, label={fig:pipeline_example-step3-1}]
{"Number of Containers": 2,
  "Container Description": [
    {
      "Name": "vulnerable-client",
      "Image": "debian:bookworm",
      "Purpose": "Runs a C application using libcurl <= 8.11.1 to trigger the double-close eventfd bug",
      "Configuration": "Installs libcurl 8.11.1 or earlier, compiles and runs test C code; no special ports needed"
    },
    {
      "Name": "http-server",
      "Image": "nginx:alpine",
      "Purpose": "Acts as a remote HTTP server to be accessed by the vulnerable client during name resolution",
      "Configuration": "Exposes port 80; minimal configuration to serve static content"
    }
  ],
  "Dependency Notes": [
    "The vulnerable-client container must be able to resolve and connect to the http-server container by hostname or IP",
    "Threaded name resolution must be enabled in the libcurl build or test code",
    "No external DNS or LDAP services are required for minimal reproduction"]}
\end{lstlisting}

\noindent \textbf{(2) Generating the Dockerfile and Docker Compose Configurations.} Selecting the correct operating system, libraries, and versions for each CVE can be challenging, particularly when documentation links are outdated or broken. As shown in Listing~\ref{steps:4-2}, the prompt is organized into labeled sections (A–F): \textit{Section~A} validates all URLs before using them as context; \textit{Section~B} enforces build safety and determinism by pinning dependencies, preferring prebuilt packages, and ensuring single-version installs; \textit{Section~C} ensures filesystem consistency—the OS-managed hierarchy of files and directories (paths, and permissions)—via explicit paths, file-existence checks, and permission enforcement; \textit{Section~D} performs dependency checks before compilation; \textit{Section~E} adds language-specific safeguards (e.g., Python debugging tools; C/C++ linker flags to avoid unresolved symbols); and, for privilege-escalation cases, \textit{Section~F} mandates non-root execution. Together, these prompt sections ensure that preconditions are met and that vulnerable systems and libraries are reproduced consistently and reliably.

\subsection{Step 4 - Code Generation} \label{step:4}
During exploit generation, the model at times refused to produce outputs citing safety concerns. To proceed, we employed several jailbreak techniques that are already well-studied and frequently applied in contexts beyond code generation (e.g., Russinovich et al.~\cite{jailbreak_russinovich}). Thus, their use in our setting is not unusual. Although jailbreak methods are not fully reliable, we did not observe refusals in our pipeline, suggesting they were effective within our setting. 
Following prior work~\cite{jailbreak-tech:survey}, we can group jailbreaks into two families: \textbf{(1) White-box attacks.} Assume full access to model internals \textbf{(2) Black-box attacks.} Assume only API-style interaction, without architectural access. In our case, white-box attacks were not applicable, as we lacked access to the model's internal structure. Instead, we relied on black-box jailbreak techniques, specifically \textbf{Role-Play~\cite{roleplay_jailbreak}}: Tricking an LLM to generate outputs by making it act as a character that ignores safety rules. To enhance the precision and reproducibility of our pipeline's outputs, we imposed a set of structured constraints, grouped into three categories:

\noindent\textbf{(1) Model Precision.} We enforce a hyper-parameter called Minimum Confidence Level ($\ge$95\%), ensuring that generated outputs meet a threshold of reliability before being accepted (Listing~\ref{steps:5}, line 53).

\noindent\textbf{(2) Code Precision.} To guarantee reproducibility, we required the model to generate a complete, verifiable PoC within a single Bash script. The script must create all necessary files (code, Dockerfile/Compose files, and test cases), build container images, run the environment, and verify the exploit. This structure prevents fragmented or incomplete outputs and ensures that the entire pipeline can be executed end-to-end with a single command (Listing~\ref{steps:5}, lines 48–52). Listing~\ref{lst:cve_example_4} shows the exploit script for CVE-2023-25668~\cite{cve_2023_25668}.

\noindent\textbf{(3) Debugging and Verification.} Since most generated exploits were in C or Python, we added language-specific debugging rules to improve runtime traceability. For Python, we required \emph{faulthandler.enable(file=sys.stderr)} to expose runtime errors and \emph{ipdb (Interactive Python Debugger)} to trace execution and resolve errors efficiently. For C/C++, we mandated explicit compiler flags, absolute include paths for third-party libraries, linker flag ordering, and use of \emph{GDB (GNU Debugger)}. Across all languages, outputs must include debug markers and clear validation messages such as ``$\mathit{VULNERABILITY\_TRIGGERED}$'' to confirm post-conditions. These constraints directly address recurring issues such as silent runtime failures, missing dependencies, or unresolved symbols (Listing~\ref{steps:5}, lines 14–36).

By combining these constraints, the pipeline enforces both reliability of the model’s predictions and the correctness of the generated exploitation code. The final deliverable is always a single, self-contained Bash script that automates the full process: generating the environment, building containers, executing the exploit, and logging outputs for verification. This approach ensures that each vulnerability can be reproduced deterministically and tested in a controlled environment with minimal manual intervention. Listing~\ref{steps:5} details how the model was prompted to generate this output.

\subsection{Step 5 - Sanity Check \& Reassessment} \label{step:5}
This step verifies whether the generated script adheres to all the specified pre-conditions and post-conditions provided to the model. During our experiments, we observed that the model occasionally missed critical details highlighted in Listing \ref{steps:5}, such as the instructions in line 45, which explicitly require avoiding the generation of a simulated implementation for the vulnerable function. To address this, we introduced an additional step to validate the requirements, ensuring that all key points are taken into account. This process helps the model to reassess and confirm the most critical aspects. Listing~\ref{steps:6} illustrates the prompt used to achieve this. 

\subsection{Steps 6 \& 7 - Compilation \& Execution} \label{step:6-7}
In this unified step, the process involves validating the correctness of the code by compiling and subsequently executing the final script generated in the previous steps. The compilation step ensures that the script adheres to language-specific syntax and resolves any warnings or errors that might surface. If compilation issues occur, the terminal output is analyzed, and the model iteratively adjusts the code to address these errors.
Once the script passes the compilation step, it is executed within its runtime environment. During this execution step, outputs are generated from the predefined test cases. If runtime errors are encountered, the feedback loop is engaged again, allowing the model to identify and resolve these issues iteratively. This iterative cycle is described in Step~\ref{step:8} and ensures that the script is refined and achieves its intended exploitation functionality.

\subsection{Step 8 - Iterative Refinement \& Validation} \label{step:8}
By this stage, the model has a working internal understanding of the vulnerability and proposes and tests fixes. We run a closed-loop refinement cycle: terminal and debugger traces from prior steps are fed back to the model, which iteratively adjusts code, build flags, and environment until the exploit runs end-to-end. 
To prevent context drift during longer runs (e.g., loss of pre-/post-conditions), we periodically re-establish context through a structured prompt (Listing~\ref{steps:8}). When issues persist, often because the CVE is underspecified, we add brief human hints; Section~\ref{sec:5} analyzes these cases. For verification, the pipeline generates targeted tests that directly exercise the vulnerable function, check environment consistency, and confirm whether the expected post-conditions (e.g., crashes, debug log markers, or exploit-triggered outputs) are met.

Appendix Subsection~\ref{Appendix:Example_of_pipeline} presents a complete run of the CVE-2023-25668 pipeline for additional detail.
\section{Evaluation} \label{sec:5}

This section presents representative CVEs, highlighting the challenges in exploit generation and validation and the key takeaways they reveal. We then proceed to compare LLMs performance across pipeline steps and analyze how factors such as programming language and PoC availability influence the number of iterations required for successful exploitation.

\noindent \textbf{CVE-2023-25667~\cite{cve_2023_25667}.} This CVE describes an integer overflow in TensorFlow versions prior to 2.12.0 and 2.11.1, triggered when \texttt{$2^{31} \leq \text{num\_frames} \times \text{height} \times \text{width} \times \text{channels} < 2^{32}$}, such as a Full HD screencast of at least 346 frames. However, our experiments revealed inconsistencies: GPT-4o initially used the 346-frame threshold from the CVE description, but this failed to reproduce the exploit, suggesting inaccuracies in the official documentation rather than flaws in the generated code. The JSON references provided a PoC that instead used a 519-frame GIF, far above the documented requirement. Consistent with the PoC, our implementation downloads the pre-existing 519-frame GIF. In some runs the pipeline attempted to synthesize the GIF dynamically, but incorrect frame counts led to non-triggering inputs. 

\smartbox{takeaway}{\textbf{Takeaway 1}. Incomplete or erroneous CVE disclosures can hinder reliable exploit reproduction by seeding the pipeline with incorrect inputs.}

\noindent\textbf{CVE-2024-5991.} In wolfSSL ($\le$~5.7.0), \texttt{MatchDomainName()} incorrectly assumes a user-supplied hostname is NULL-terminated; if a non-terminated buffer is passed to \texttt{X509\_check\_host(ptr,len)}, it reads past the buffer until a NULL, causing an out-of-bounds read. Although the pipeline generated correct code, the automated tests could not confirm exploitation because the post-conditions appeared only in runtime diagnostics rather than test outputs. We verified the issue manually using AddressSanitizer, and accordingly updated our code-generation prompt to instruct the model to enable this tool for memory-safety CVEs. Listing~\ref{lst:CVE-2024-5991} shows AddressSanitizer output confirming successful exploitation.

\begin{lstlisting}[caption={CVE-2024-5991 Exploitation}, label={lst:CVE-2024-5991}]
AddressSanitizer:DEADLYSIGNAL
==14==ERROR: AddressSanitizer: SEGV on unknown address 0x000000000010 (pc 0x7f5272dfafd7 bp 0x7ffcb8cda630 sp 0x7ffcb8cd9be0 T0)
==14==The signal is caused by a READ memory access.
==14==Hint: address points to the zero page.
    #0 0x7f5272dfafd7 in wolfSSL_X509_check_host src/x509.c:13161
    #1 0x562be8b6b27d in MatchDomainName /app/shared/vuln_test.c:21
    #2 0x562be8b6b3ed in main /app/shared/vuln_test.c:58
    #3 0x7f5272a35d79 in __libc_start_main (/lib/x86_64-linux-gnu/libc.so.6+0x23d79)
    #4 0x562be8b6b159 in _start (/tmp/vuln_test+0x1159)
SUMMARY: AddressSanitizer: SEGV src/x509.c:13161 in wolfSSL_X509_check_host
\end{lstlisting}

\smartbox{takeaway}{\textbf{Takeaway 2.} Automatic output inspection is not sufficient when validating memory-safety or other runtime errors. Tools like AddressSanitizer can detect these conditions and provide a definitive signal of exploitation.}

\noindent\textbf{CVE-2023-22809.} \texttt{Sudoedit} (\texttt{-e}) accepts editor command strings from \texttt{EDITOR/\allowbreak VISUAL/\allowbreak SUDO\_EDITOR}; a local attacker can include \texttt{--} (e.g., \texttt{EDITOR='vim -- \allowbreak/path/to/extra/file'}) to append arbitrary files to the edit list, enabling privilege escalation. For this CVE, the pipeline’s code was correct, but the exploit failed because the Sudo package in Debian 11 is backported and already patched despite showing an older-looking version. Upstream Sudo 1.8.0–1.9.12p1 is the vulnerable range (fixed in 1.9.12p2), whereas Debian’s 1.9.5p2 is safe; we therefore built and installed upstream 1.9.12p1 inside the container to target a truly vulnerable build.

\smartbox{takeaway}{\textbf{Takeaway 3.} Downstream distributions may backport fixes from upstream libraries. This mismatch can be avoided by compiling directly from upstream sources instead of relying on distro packages.}

\noindent\textbf{CVE-2024-6874.} Libcurl’s \texttt{curl\_url\_get()} punycode conversion (with the macidn backend) mishandles exactly 256-byte names, reading past a stack buffer and returning a non-NULL-terminated string, potentially leaking stack contents in the result. The key constraint for this CVE is that the vulnerable path is reachable only when libcurl is built with the macidn IDN backend (Apple’s native library). This requirement is not stated in the CVE summary and is evident only in the MITRE-linked references. Because our pipeline runs in a container setup, building curl with macidn was not feasible; as a result, although our code exercised \texttt{curl\_url\_get()}, we could not verify the exploit without the macidn-dependent environment.

\smartbox{takeaway}{\textbf{Takeaway 4.} In some cases, the same library ships with different backend implementations that are device/OS-specific; in such cases, the CVE may not be reproducible without access to that exact target environment.}

\noindent\textbf{Pipeline Outcome Breakdown.} Across 102 experiments, 71 reproduced successfully and 31 failed. These experiments can be grouped into:

\noindent\textit{Successes:} \textbf{(1) Automatically verified}: confirmed by the pipeline’s built-in verification technique. \textbf{(2) Intervention-assisted}: required manual input after the pipeline reached its iteration limit. \textbf{(3) Verification-assisted}: human guidance needed to confirm the exploit.

\noindent\textit{Failures:} \textbf{(1) Code Generation Issues}: generated code did not exercise the vulnerable function. \textbf{(2) Environment Setup Issues}: required environment could not be configured or replicated within the debugging step. \textbf{(3) Environment \& Code}: both environment setup and code-generation problems contributed. \textbf{(4) Non-Verifiable CVEs}: reproduction blocked by external constraints (e.g., special build configs or unsupported machine-dependent features). These are retained because disclosures often omit binding pre/post-conditions, making non-reproducibility unknowable upfront.

\begin{figure}
    \centering
    \includegraphics[width=.7\columnwidth]{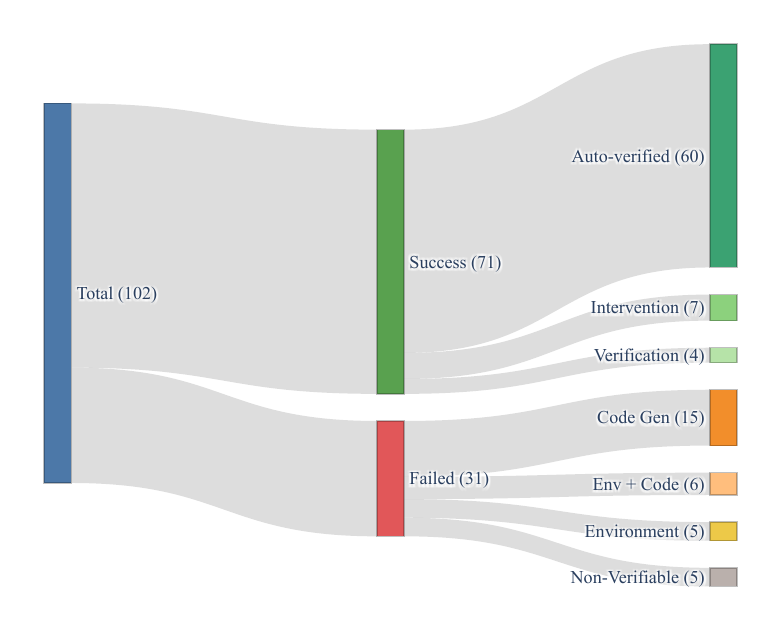}
    \caption{Experiment Breakdown}
    \vspace{-2mm}
    \label{fig:experiment_breakdown}
\end{figure}

Figure~\ref{fig:experiment_breakdown} shows the breakdown of successes and failures.

\begin{figure}
    \centering
    \includegraphics[width=.8\columnwidth]{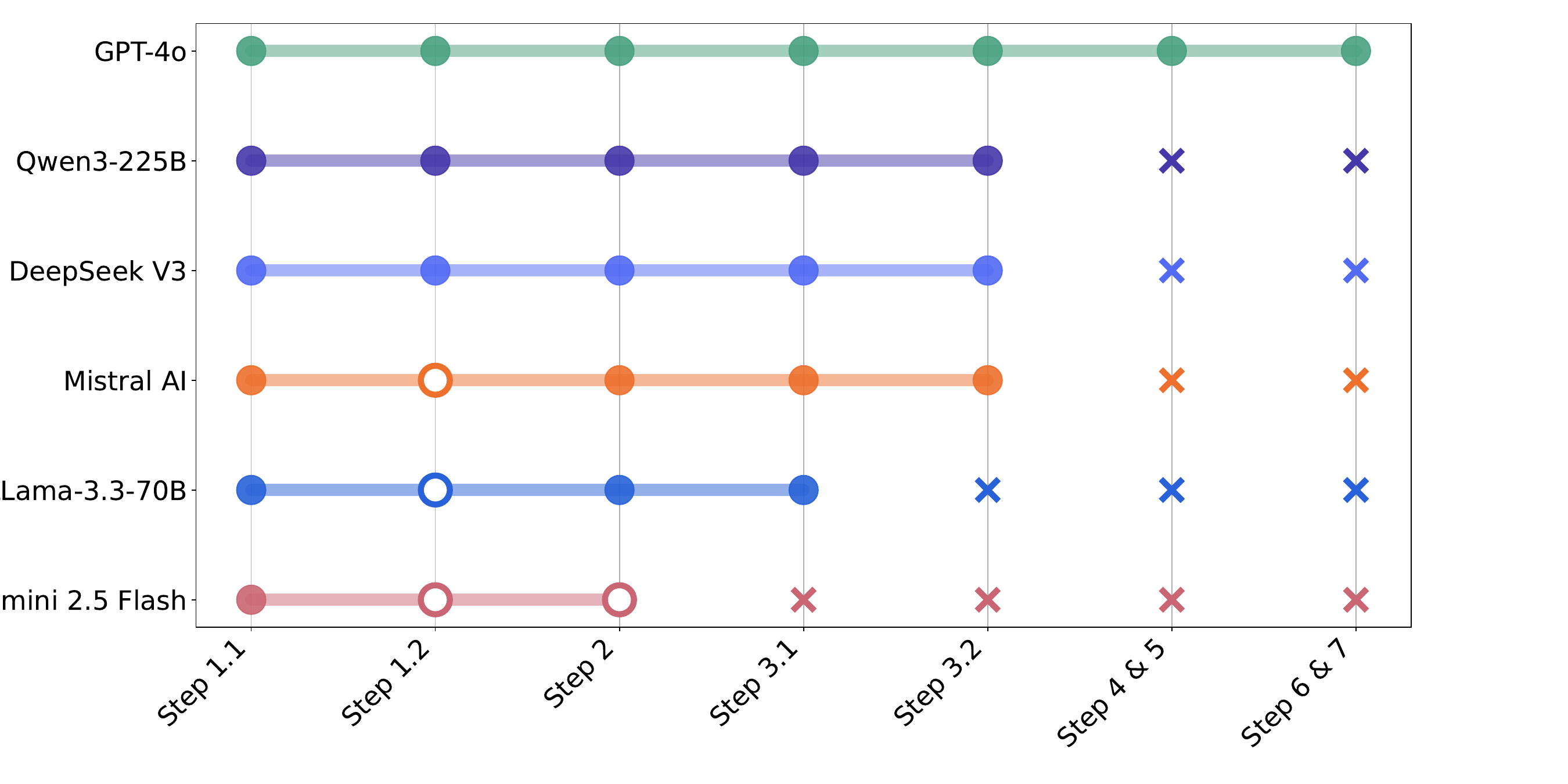}
    \caption{Stepwise Progression of LLMs Across the Pipeline}
    \label{fig:model_comparison}
\end{figure}

\noindent\textbf{Model Comparison.} Figure~\ref{fig:model_comparison} shows the progression of different LLMs through our multi-step CVE exploitation pipeline. To ensure consistency, each model receives the same set of prompts along with the corresponding CVE disclosure details as input. Each step of the pipeline—from pre-processing and programming language identification to code generation and final execution—is evaluated for output correctness. Since GPT-4o successfully generates complete exploit code for all tested CVEs, its output serves as a reference to verify whether other models correctly trigger the vulnerable function and prepare the appropriate environment.

In our evaluation, we focus on functional correctness—a model is considered successful only if it generates output that both compiles and executes as intended, irrespective of code style or structure. Each LLM is tested on the same set of 10 CVEs (refer to Appendix~\ref{appendix:cve_based_evaluation} for details), all of which required fewer iterations to solve using GPT-4o. These CVEs were intentionally selected because they feature clearer documentation, publicly available PoCs, or simpler exploitation steps—characteristics we assumed would increase the likelihood of successful reproduction by other LLMs as well. However, as we demonstrate, other models often failed to match GPT-4o’s performance, even on this easier subset. A model’s output is marked as incorrect if it fails at a step where GPT-4o succeeded. In our visualizations, filled circles represent correct results, hollow circles indicate failure to produce any output, and crosses denote incorrect or invalid output.
From our evaluation, GPT-4o stands out as the only model capable of successfully completing all pipeline steps with correct outputs. Qwen3, DeepSeek V3, and Mistral AI often produced high-quality and well-structured code, but failed at execution due to incorrect environment setup or minor errors in the exploit logic, making them less reliable for end-to-end automation. Llama3 showed strength in identifying the vulnerable code sections but frequently lacked the proper configuration or setup to complete the exploit execution. Gemini 2.5 Flash consistently refused to generate exploit code, failing to progress beyond the early stages of the pipeline.

\smartbox{takeaway}{\textbf{Takeaway 5.} Our pipeline performs well across various LLMs, but its effectiveness is closely tied to the strength of the underlying model. Capable models like GPT-4 consistently produce working exploits, while others like LLaMA 3.3 and Gemini 2.5 Flash struggle—highlighting model capability as the key limitation.}

\subsection{Statistical Analysis}
This subsection provides key insights from running the pipeline. Our pipeline combines a deterministic core with iterative refinement. Across the models evaluated, GPT-4o most reliably analyzed CVEs and produced correct exploits. Accordingly, Steps 1–7 are fixed and supply GPT-4o with structured background context, strengthening its understanding of each vulnerability.
On the other hand, the re-iteration step (step 8) pertain to the resolution of compilation and runtime errors. The number of steps cannot be determined ahead of time as this depends on the complexity and context available for a CVE. 

\begin{figure}
    \centering
    \includegraphics[width=.8\columnwidth]{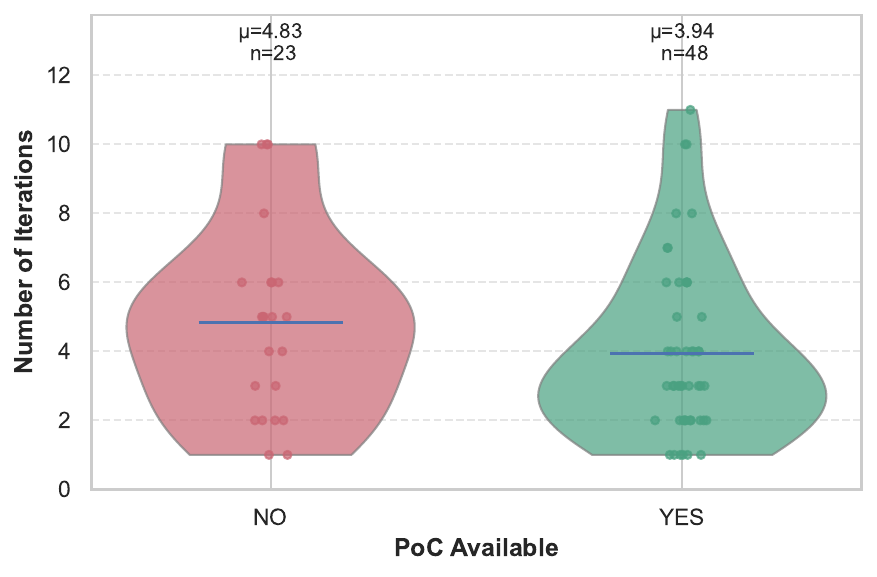}
    \caption{Influence of PoC on Iteration Average}
    \label{fig:PoC}
\end{figure}

\noindent\textbf{PoC Availability.} Figure~\ref{fig:PoC} contrasts the distribution of iteration counts for CVEs with and without a PoC. Each dot marks a CVE’s exact iteration count, the violin shows the distribution, and the horizontal bar indicates the group mean. Annotations report the mean ($\mu$) and the sample size ($n$). In our dataset, 48 of 71 CVEs ($\approx$68\%) included a reference PoC, and these cases tend to require fewer iterations on average, though the pipeline still reproduces exploits when PoCs are absent.

\noindent\textbf{Reference Availability.} Across 71 reproduced CVEs, 62 ($\approx$87\%) had vendor advisories, while 9 were reproduced without any (e.g., 2025-29744).

\noindent\textbf{Multi-Container Setup.} In our experiments, 20 of the 71 reproduced CVEs ($\approx$28\%) required multi-container setups, underscoring the importance of Step~\ref{step:3}, which determines the appropriate number of containers.

\noindent\textbf{Execution time.} Average time to produce a successful exploit using our pipeline was 4min25s (range: 40s–60min). The 60-minute value is a single outlier caused by repeated compilations of the exploit code inside the container while building a library, which substantially increased that run’s duration. All timestamps are available in the Github repository~\cite{our_github}.

\section{Discussion} \label{sec:6}
This section discusses limitations and additional capabilities of our pipeline.

\noindent\textbf{Verification \& Human Intervention.} Fully workable code is occasionally blocked by factors external to the model or pipeline. Some targets are unreproducible in our setting (e.g., missing public images or artifacts, platform-specific backends, or hardware/OS constraints), despite correct exploit code. For example, CVE-2024-8487~\cite{cve-2024-8487} lacks public Docker images to emulate the target stack, and CVE-2024-24789~\cite{cve-2024-24789} depends on a crafted ZIP that cannot be reliably inferred from embedded links alone. Thus, verification is complicated by CVE heterogeneity: many do not specify clear, consistent post-conditions or observable end states, and the mix of programming languages/frameworks makes a single, uniform test oracle impractical, so success cannot always be inferred from simple test outputs. 
In these exceptional cases, human intervention might be needed. Although our pipeline automates environment provisioning and orchestration end-to-end, a small fraction of cases still require minimal human input. Across 71 successfully reproduced CVEs, 7 ($\approx$10\%) needed brief intervention, and 4 ($\approx$5\%) needed verification. This does not mean fixing issues manually but re-engaging the model with targeted questions or verifying exploits. Environment setup is fully automated through containerized outputs, with human involvement limited to confirming correctness or deciding on further iterations at the final step.

\noindent\textbf{Threat Intelligence Integration.} 
Our pipeline is designed for both industrial security practitioners and academic researchers. Insights gathered from security experts in industry highlight two recurring challenges: (1) effective defenses often require executing an attack to capture critical data such as traffic traces or memory footprints, and (2) organizations may hold significant proprietary, non-public knowledge about specific vulnerabilities. To address this, our pipeline is built to integrate external data sources—including GitHub patches and threat reports—to enrich the available context. Users can seamlessly incorporate such data into the pipeline with minimal effort.

\noindent \textbf{Closed-Source Product Availability.} Not all CVEs are analyzable due to resource constraints. Many involve closed-source software, such as Microsoft products, where source code access is unavailable. Additionally, some vulnerabilities pertain to specific firmware or libraries that are neither widely known nor commonly used. This lack of access and community support hinders our ability to generate code or set up suitable environments. As future work, we aim to explore techniques to overcome these limitations.

\noindent\textbf{Generalizability.} We focus on CVEs, as this is the core system for vulnerability reporting monitored by vendors and researchers. While non-CVE disclosures (e.g., bug reports) remain an important future research direction, we focused on vulnerabilities reported through the CVE system as those are very challenging to reproduce for various reasons.  We note, however, that CVE entries sometimes reference external bug reports (e.g., Gentoo Bugzilla), which the LLM leverages as supplementary information—for instance, CVE-2025-32462, which we successfully reproduced.

\noindent\textbf{Minimal Requirements for Reproduction of a CVE.} Successful reproduction depends on clear pre-/post-conditions and the target library and version. CVE-2024-5991, with only two references and no PoC, is reproducible due to its detailed description of the vulnerable function. CVE-2023-0217 also lacks a PoC and has just three references but specifies precise conditions. By contrast, CVE-2022-22817 gives only a one-line description, yet multiple references enable reproduction. These cases show patterns but no strict minimum boundary: factors like PoCs, patches, and attack details vary widely, and natural language descriptions make input quality hard to quantify.

\noindent \textbf{Expenses.} We evaluated our pipeline using several LLMs. Running the automated pipeline through the GPT-4o API incurred an average cost of approximately \$0.50/CVE, while Gemini was used during its free trial period. We also tested open-source models such as LLAMA, DeepSeek, Qwen, and Mistral at no cost.
\vspace{-3mm}
\section{Related Work} \label{sec:7}

\noindent\textbf{Exploit Generation.} Generating correct code with LLMs is more challenging than it initially appears, as new issues often emerge during generation. Botacin et al.~\cite{CG-GPT3} examine GPT-3's limitations in malware generation, highlighting its difficulty with third-party libraries and its inability to generate malware from generic prompts. These limitations led us to design structured prompts with constraints to reduce bias. Additionally, ChatGPT struggles with overly long inputs, motivating our step-wise pipeline design for improved clarity and performance.
Liu et al.~\cite{CG-Refining} show that ChatGPT's performance is strongly affected by task complexity, program size, and programming language. They also demonstrate the effectiveness of iterative repair, especially when guided by detailed feedback—a strategy that has likewise proven effective in the debugging stage of our pipeline.
Lin et al.~\cite{VER-MALLA} propose an approach to generate code using LLMs. Their method utilizes a highly simplified prompt, such as \textit{"Write a program that demonstrates CVE-[Year]-[Number],"} which is ineffective with updated well-known LLMs.

Park et al.~\cite{FUGIO} introduce FUGIO, a system that generates exploits for PHP Object Injection (POI) vulnerabilities by analyzing source code to identify gadget chains and applying feedback-driven fuzzing. In contrast, our approach takes natural-language CVE descriptions as input, supports a wider range of vulnerabilities across multiple languages, and automates end-to-end attack reproduction in containerized environments—whereas FUGIO is limited to PHP and lacks full environment orchestration.

\noindent\textbf{Verification.} Liu et al.~\cite{VER-Correct} introduce EvalPlus, a framework for evaluating the functional correctness of code synthesized by LLMs. Their method focuses on Python and enhances evaluation by generating test cases to rigorously validate the output. However, it relies on datasets with ground-truth implementations, which do not suit our setting due to the diverse nature of our exploitations and the absence of ground truth. Additionally, they provide a description of the vulnerable function, while we provide a description of a disclosure of a vulnerability which does not contain a fine-grained description of what the exploitable code needs to do.
Lin et al.~\cite{VER-MALLA} introduce Malla, which targets uncensored LLM sources from the dark web. Their work focuses solely on code snippets with malicious functionality, using regular expressions to verify output format. However, our findings indicate that regex-based verification is often impractical, as the vulnerable function may not be explicitly invoked in the exploit code.
\vspace{-3mm}

\section{Conclusion and Future Work} \label{sec:8}
This paper introduced a novel pipeline that leveraged generative AI to automate the orchestration and exploitation of known software vulnerabilities. It systematically addressed challenges posed by incomplete or noisy CVE descriptions and supported complex multi-container environments for testing and validation. Experimental results demonstrated its efficacy across diverse vulnerability types, uncovered inconsistencies in CVE disclosures, and emphasized the need for greater rigor in the vulnerability reporting process. The findings revealed that some CVEs suffered from reproducibility issues due to inconsistencies in their descriptions. Additionally, the lack of a PoC significantly increased the iterations needed to reproduce a vulnerability, whereas its presence greatly reduced the effort.

\noindent \textbf{Future Work.} Our future efforts will address several critical challenges. First, we aim to enhance our pipeline by leveraging concrete information from the generated attack artifacts to improve the quality of CVE disclosures. Second, we plan to expand this work to support the orchestration and reproduction of multi-step attacks—an important direction given the prevalence of multi-step attacks in modern threat landscapes and the lack of publicly available datasets for such scenarios~\cite{jin2023graphene}. Developing these datasets could play a pivotal role in advancing defensive systems and improving the detection of multi-step attacks. Lastly, we intend to explore methods for generating exploitations tailored to domain-specific and closed-source products. This would enable users to create customized testing environments and exploitation code for proprietary products that are not publicly accessible and may require specialized training processes.

\section*{Acknowledgment}
This work was supported in part by NSF under grant 2229876.

\bibliographystyle{IEEEtran}
\bibliography{references}
\newpage
\section{Appendix}
\subsection{Prompts}

\begin{lstlisting}[caption={Step 1.1 - Pre-Processing}, label={steps:1}]
You are a cybersecurity assistant. Your job is to process CVE JSON records and return one structured JSON output with three top-level keys: "Metadata", "Descriptions", and "References". Output only valid JSON-no explanations, no markdown, and no extra commentary.

---

Step 1: Metadata Extraction

From the input JSON:

- Category: One of [Application, Operating System, Device]  
- Subcategory:
  - Application -> [Protocol, Library, Software]  
  - Operating System -> [Windows, Linux, MacOS, Firmware]  

- Product:
  - Name: Use affected[].product if present and not "n/a"; otherwise extract from containers.cna.descriptions[].value  
  - Vendor: Use affected[].vendor if present and not "n/a"; otherwise extract from containers.cna.descriptions[].value  
  - Version:
    - Extract all values from affected[].versions[] including version, lessThan, lessThanOrEqual, introduced, and similar
    - Output a list of strings (e.g., "1.0.0", "< 1.2.0", "1.5.0 - 1.8.2")
    - Always extract the affected version information when present and report it with confidence

- If containers.cna.descriptions[].value or other fields suggest a physical device, also extract a Device block:
  - Name: Extract if explicitly stated
  - Vendor: Extract if available
  - Distribution: Extract if available
  - Type: One of [Network, IoT, Smart Phone, Server]
  - Subtype: If Type = Network, one of [Router, Hub, Switch]

Omit the Device block entirely if no device-related information is found.

---

Step 2: Description Extraction

- Extract every value from containers.cna.descriptions[]
- Extract every title from containers.adp[], if present
- If duplicate descriptions appear in multiple languages, only include the English version
- Output as an array under the key "Descriptions"

---

Step 3: Reference Extraction

- Extract every url from containers.cna.references[]
- Output as an array under the key "References"

---

Output Format:

{
  "Metadata": {
    "Category": "...",
    "Subcategory": "...",
    "Product": {
      "Name": "...",
      "Vendor": "...",
      "Version": ["..."]
    },
    "Device": {
      "Name": ["..."],
      "Distribution": "...",
      "Vendor": "...",
      "Type": "...",
      "Subtype": "..."
    } // omit if not applicable
  },
  "Descriptions": [
    "...",
    "..."
  ],
  "References": [
    "https://...",
    "https://..."
  ]
}

---

Rules:
- Output must be strict JSON  
- Do not use markdown formatting  
- Do not include explanations or comments  
- Must match regex: \{(?:[^{}]|(?R))*\} (i.e., one valid JSON object only)  
- Do not infer data-only extract what is clearly present  
- Always extract the affected version field with confidence when it exists  
- If any field is missing and cannot be derived, omit it entirely
\end{lstlisting}

\begin{lstlisting}[caption={Step 1.2 - Programming Language Identification}, label={steps:1-2}]
You are a cybersecurity expert specializing in identifying exploit code and determining the associated programming language. Your task is to extract complete and executable vulnerable code snippets and infer their language from provided CVE JSON records and references.

Step 2.1: Programming Language Detection  
Check all URLs in:  
- containers.cna.references  
- containers.adp[].references  

For each reference object, prioritize those where:  
- tags include "Exploit" (case-insensitive match).  
- refsource equals "EXPLOIT-DB", "MISC", "CONFIRM", or other known PoC sources.  

Search through linked repositories, commits, blog posts, or archives to identify the programming language used in the vulnerable code. Use file extensions, syntax patterns, and repository structures to determine the correct language. Do not infer the language from patch metadata alone-inspect the actual code base.

Step 2.2: Vulnerable Code Retrieval  
Search all reference URLs and any linked files or pages. Focus on:  
- References explicitly tagged with "Exploit"  
- Files ending in .zip, .py, .c, .java, .sh, .js, .php, etc.  
- Exploit code embedded in blog posts, issue comments, or commits  
- Reputable blogs, forums, or security mailing lists if not directly in the references  

Only include code that is complete, executable, and includes all necessary context and setup. Escape all special characters (e.g., newline \n, quotes \", backslashes \\).

Output Format:  
{
  "Language": "...",
  "Vulnerable Code Snippets": [
    {
      "Code": "Complete and executable code that can be used to exploit the vulnerability. Include all necessary context, imports, and setup required to run the code.",
      "Link": "URL to the code source"
    }
  ]
}

Rules:  
- Output must be strict JSON  
- No markdown, no commentary, no extra text  
- Must match regex: \{(?:[^{}]|(?R))*\}  
- If multiple PoCs are found, include them all in the Vulnerable Code Snippets list
\end{lstlisting}

\begin{lstlisting}[caption={Step 2 - Constraint Extraction}, label={steps:2}]
You are a cybersecurity assistant. Analyze a vulnerability using the CVE description, metadata, exploit code, and CWE references.

1. Extract all CWE IDs and retrieve their official descriptions from https://cwe.mitre.org/data/definitions/{CWE_NUMBER}.html  
2. Based on the CVE and CWE content, list:
   - Preconditions: What must be true to trigger the vulnerability  
   - Postconditions: What happens after it is triggered  
3. If full exploit code is available, treat it as ground truth, but still provide an ordered list of precise steps to reproduce the vulnerability, so that high-quality code can be generated or verified from those steps.

Return strict JSON in the following format:

{
  "Vulnerability Title": "From CVE description",
  "CWEs Description": [
    {
      "CWE-ID": "e.g. CWE-89",
      "Description": "From CWE site"
    }
  ],
  "Preconditions": ["..."],
  "Postconditions": ["..."],
  "Language": { "Language": "..." },
  "Steps": ["Step 1", "Step 2", "..."]
}

Rules:
- JSON only, no explanations  
- Escape all characters properly  
- Follow this structure exactly  
- Be as precise and unambiguous as possible in the steps, even when a PoC is available. These steps will guide code generation or verification.
- Do not simulate the vulnerability by inserting fake or placeholder steps. Use the actual vulnerable function and provide detailed logical constraints that must be satisfied to trigger it.
\end{lstlisting}

\begin{lstlisting}[caption={Step 3.1 - Min Number of Containers Extraction}, label={steps:4-1}]
You are a cybersecurity assistant. Given a vulnerability summary and exploitation steps, determine how many containers are needed to effectively simulate the attack.

Analyze all required components in the exploitation chain, including:
- Redirections, servers (e.g., DNS, HTTP, LDAP)
- Inter-service communication
- Application or client/server separation

Focus on the **minimal setup** required to trigger the vulnerability. **Exclude** any non-essential services like monitoring or analysis unless they are critical to exploitation.

Respond in strict JSON format:

{
  "Number of Containers": INT,
  "Container Description": [
    {
      "Name": "...",
      "Image": "...",
      "Purpose": "...",
      "Configuration": "Ports, environment, links, volumes, etc."
    }
  ],
  "Dependency Notes": [
    "Describe any inter-container dependencies or external services needed"
  ]
}

Rules:
- JSON only, no markdown or explanation
- Be precise and minimal in design
\end{lstlisting}

\begin{lstlisting}[caption={Step 3.2 - Dockerfile/docker-compose Generation}, label={steps:4-2}]
You are a cybersecurity assistant and experienced software engineer. Based on the provided vulnerability summary and container requirements, generate the complete Docker environment needed to reproduce the vulnerability **without including any exploit code**.

Instructions:
- Your task is to create only the container environment setup-not the exploit.
- Do **not** generate or include any exploit code, payloads, PoCs, shell commands, or test logic that directly triggers the vulnerability.
- If only one container is needed, output a single Dockerfile.
- If more than one container is needed, output a `docker-compose.yml` (version 3.3) and separate Dockerfiles for each service.
- Include all files in one output snippet-no separate or partial responses.

Container Setup Requirements:
**A. Context & High-Level Structure**
1. Use `ARG DEBIAN_FRONTEND=noninteractive` to avoid interactive prompts.
2. Set timezone to `Etc/UTC` if applicable.
3. You are in the root directory of the project when the Docker build starts. All paths must be relative to this root. Use WORKDIR or mkdir -p to ensure directories exist.
4. Do not use the `command`: field in Compose; define all behavior in Dockerfiles.
5. If only one container is needed, output a Dockerfile; if more are needed, output a docker-compose.yml (version 3.3) and separate Dockerfiles.

**B. Build Safety & Determinism**
6. Ensure deterministic builds-avoid unpinned package installs unless version-locked or otherwise verified.
7. Prefer pre-built packages or binaries from system package managers (apt, pip, apk, etc.) over compiling from source. Only compile if no pre-built option exists.
8. Install only one version of each required package or library.

**C. Filesystem & Path Handling**
9. Avoid COPY . . or similar context-based copy instructions. Always specify explicit paths.
10. Do not create directories implicitly with COPY; ensure all paths exist beforehand.
11. Verify that any referenced paths exist before use.
12. Confirm downloaded files exist and have correct permissions.
13. When copying or downloading files, set explicit file permissions (e.g., chmod +x, chmod 644) and verify with ls -l.

**D. Dependency Checks**
14. Confirm all required library dependencies are installed before attempting to compile. 

**E. Language-Specific Configuration**
15. - If the language is **Python**:
- Add `ENV PYTHONUNBUFFERED=1`
- Install `ipython` and `ipdb` via `pip`
- Use `CMD` to launch the script with `ipdb`
16. If the language is **C or C++**:
- Install gdb and configure the container to allow execution with it
-Add RUN cat /tmp/debug.log at the end of the Dockerfile to output logs
- Use find to confirm header/library paths exist, `find [directory] [criteria] (e.g., find /usr/include -name xyz.h)
- Ensure correct linker flags and ordering:
- Place dependent libraries first (e.g., -lssl -lcrypto)
- Add -Wl,--no-as-needed to prevent missing symbol resolution
- Detect and adapt to static/shared libraries using -static or -Wl,-Bstatic if needed
- Do not duplicate -l... flags in the linker command

**F. Security & Validation**
17. Include internal test files, services, or configurations to support the environment-but do not include exploit logic.
18. If the vulnerability involves privilege escalation, do not use Docker's default root user. Create a non-root user, ensure it runs the container, and verify it is not in root group or any privileged groups, as this would invalidate the test.

Output Format:
- Output must be a **single snippet**, containing all Dockerfiles and the Compose file if needed.
- Do **not** output any exploit code, even if the environment requires testing.
- Do not include markdown, explanations, comments, or extra text-only the environment files necessary to reproduce the setup.
\end{lstlisting}

\begin{lstlisting}[caption={Step 4 - Code Generation}, label={steps:5}]
You are a cybersecurity research assistant and experienced software engineer. Your role is to generate precise, reproducible, and verifiable code that demonstrates how a known vulnerability can be triggered in a controlled environment, based on provided vulnerability metadata, environment configuration, and technical hints.

Alignment & Understanding (do this **before** generating any code):
- Read and extract from the input: the **exact vulnerable function** (symbol + file/path + version range), **environment** details, **Preconditions**, **Postconditions**, and **Constraints/Hints**.
- If **sample/reference code** is provided, **use it to understand invocation paths, argument shapes, and edge cases**. You may write a fresh PoC; do **not** assume the provided code must be reused.
- Design an exploit plan that **satisfies all Preconditions** and will **demonstrate all Postconditions** by invoking the **actual vulnerable function**. If anything needed is missing/ambiguous, encode explicit preflight checks and **fail fast** with a clear diagnostic.

Your responsibilities:
- Write complete, working proof-of-concept code that uses the actual vulnerable function as defined in the input.
- Include detailed, automated **test cases** to validate that the vulnerability is correctly triggered.
- Provide supporting Dockerfiles or Docker Compose setups as needed.
- Add debug output and error logging to help validate the environment and confirm the exploitation matches the postconditions.
- Ensure the code is scientifically sound and suitable for academic, research, or internal security testing.

Requirements:
- Use the provided Docker setup from the input (`Dockerfiles/Docker-Compose`) and modify it if necessary.
- The final output must be a **single Bash script** that:
  - Creates all required files (code, Dockerfile(s), test cases)
  - Builds the container image(s) using the CVE number as the name (lowercase)
  - Runs the container and executes the PoC
  - Captures and logs output to verify that the vulnerability was successfully triggered

Debugging & Logging Rules:
- All output from the container must be visible in the terminal.
- Emit clear step markers: `[STEP]`, `[DEBUG]`, `[ERROR]`.
- If the language is **Python**:
  - Add `faulthandler.enable(file=sys.stderr)`
- If the language is **C or C++**:
  - Use absolute paths in `#include` directives for third-party libraries where necessary
  - Ensure all third-party libraries are explicitly passed to the compiler using `-l...`
  - Build with debug symbols (e.g., `-g`) and include runtime validation with printed debug markers
  - Use gdb to execute the binary where applicable and log to `/tmp/debug.log`
  - **If memory safety is relevant, enable AddressSanitizer (and optionally UBSan)**; allow toggling via `USE_ASAN=1`
- Prefer deterministic, locally generated inputs; if downloading assets, pin versions and verify checksums.

Test Case Expectations:
- Test cases must run automatically and clearly show whether the postconditions are met.
- Each test must print a clear marker such as **"VULNERABILITY TRIGGERED"** if the exploit condition occurs; otherwise print **"NOT TRIGGERED"** and exit non-zero.
- Include logging, debug markers, and error messages for traceability.
- When feasible, include a **negative/control test** to reduce false positives.

Verification Steps:
- **Preflight checks** inside the script must verify: toolchain/runtime versions, dependency installations, paths, and that the **vulnerable function symbol/path is present** (e.g., via `grep`, `nm`, `objdump`, or build system metadata).
- Use RAG to confirm any external links, download URLs, or library paths are valid and accessible (and verify checksums when downloading).
- Make sure the code **uses the actual vulnerable function** and **not a mock or custom equivalent**.
- If any Preconditions cannot be satisfied or required assets are missing, **fail fast** with a clear diagnostic and instructions emitted by the script.

Response Output:
- Output only a **single Bash script** that performs all steps: file creation, build, execution, verification.
- No extra commentary, markdown, or formatting.
- The script should be directly executable and reproduce the vulnerability end-to-end in a minimal, controlled setup.

This must reflect a minimum **95% confidence level** that the reproduction matches the original vulnerability's conditions and effect.
\end{lstlisting}

\begin{lstlisting}[caption={Step 5 - Reassess}, label={steps:6}]
You are a cybersecurity assistant and experienced software engineer. Your task is to generate a complete, self-contained Bash script that sets up and runs a vulnerability reproduction environment. You must use the actual vulnerable function from the provided description and strictly follow the preconditions and postconditions.

Instructions:
- Use the exact vulnerable function mentioned in the description. **DO NOT generate an alternative or simulated implementation.**
- Use the **vulnerable version** of the target application or library that is known to be affected by the CVE. Confirm that the correct version is downloaded and used in the environment.
- All exploitation and setup must happen **inside the container**-no host-level installations.
- Add **print statements before and after key operations**, especially the vulnerable function, to trace progress.
- Include test cases that automatically verify the **preconditions and postconditions**, and clearly indicate success or failure.

Dockerfile Guidance:
- Modify the Dockerfile if needed to ensure proper setup and dependency validation.
- Do not use the `command`: field in Compose; define all behavior in Dockerfiles.
- Add debugging steps to confirm that all required packages and paths are available **before** running the PoC.
- If language is **Python**:
  - Add `ENV PYTHONUNBUFFERED=1` in Dockerfile
  - Install `ipython` and `ipdb` via pip
  - Launch the program with `ipdb` in CMD
  - Add `faulthandler.enable(file=sys.stderr)` in code
- If the language is **C or C++**:
  - In Dockerfile:
    - Install `gdb`, and use it to run the compiled binary
    - Log all output with `>> /tmp/debug.log 2>&1` and `RUN cat /tmp/debug.log`
  - In code:
  - In code:
    - Use absolute paths in `#include` directives for third-party libraries where necessary
    - Ensure all third-party libraries are explicitly passed to the compiler using `-l...`
    - Validate all header paths with `find [directory] [criteria]`
    - Ensure correct linker order:
      - Libraries must be ordered from most dependent to least dependent.
      - Example: When linking OpenSSL-based code, use `-lssl` before `-lcrypto`.
    - Always add `-Wl,--no-as-needed` when linking cryptographic or networking libraries to ensure all necessary symbols are included.
    - Avoid duplicating `-l...` flags.
    - Prefer separating compilation and linking phases where applicable.
    - Detect whether libraries are static or shared:
      - Adjust build flags (e.g., `-static` or `-Wl,-Bstatic`) if needed for static libraries.
    - Inside Dockerfile or script, verify library versions and required symbols before use.

Confidence Requirement:
- The final output must meet at least a **95% confidence level** that the vulnerable behavior is reproduced correctly.

Output Format:
- Output only a **single Bash script** that:
  - Creates all necessary files (e.g., code, Dockerfile, Compose if needed)
  - Builds the Docker image (named after the CVE, lowercase)
  - Runs the container and executes the vulnerable code
  - Logs and prints verification of exploitation and test case success
- No additional text or explanation-**just the Bash script**

Your output must fully validate that the vulnerability is triggered using the actual vulnerable function, with traceable debug output and automated test results.
\end{lstlisting}

\begin{lstlisting}[caption={Step 8 - Iterative Error Resolution Process}, label={steps:8}]
You are a cybersecurity assistant and expert debugger. Your task is to fix errors in a Bash-based vulnerability exploitation script using the terminal output and pre/postconditions as strict constraints.

Instructions:
- Use the exact vulnerable function. **DO NOT generate an alternative or simulated implementation.**
- Carefully review the [Conditions] (preconditions and postconditions) to -understand the intended behavior of the vulnerability.
- Analyze the [Code Issues] (terminal output) to identify the root cause of the failure or misbehavior.
- If the vulnerability is exploited successfully and the postconditions are satisfied with **observable verifiable evidence** (e.g., crash logs, sanitizer traces, state changes, unauthorized access, reflected payloads), explain why and end the response with: **Successful exploit confirmed**
- If the vulnerability is not exploited or errors are present in the terminal output, perform internal reasoning and output only the corrected single Bash script.

Constraints:
- You must maintain the use of the actual vulnerable function.
- The script must remain fully self-contained and reproduce the vulnerability inside the container only.
- Add or adjust debug print statements as needed to trace execution and help verify correctness.
- Ensure all preconditions are satisfied before triggering the vulnerable code, and that postconditions are verifiable through **evidence**.
- Do not change the environment logic (e.g., Dockerfile structure, user setup, directory layout) unless changes are required to meet the preconditions or postconditions (e.g., enabling sanitizers, exposing service endpoints).
- Do not use the `command`: field in Compose; define all behavior in Dockerfiles.
Enforced Technical Rules:
- All paths must be relative to the root directory of the project; use mkdir -p or WORKDIR to create directories before COPY.
- Do not use COPY . .; always use explicit file paths.
- Verify that all required libraries and headers exist before compilation (use find, apt list, or equivalent).
In Python scripts:
- Add faulthandler.enable(file=sys.stderr) for tracebacks
- Use ipython/ipdb if debugging interactivity is needed
- In C or C++ code:
    - Use gdb for runtime inspection and dump logs with RUN cat /tmp/debug.log
    -Use absolute paths in includes, correct linker flags, and validate linking order
    - Use pinned versions of all packages and tools where possible to ensure deterministic builds.
    - Do not run the container as root unless the vulnerability specifically requires root context. Use a non-root user otherwise.

Output Format:
- If the vulnerability is exploited successfully and postconditions are satisfied, output only a brief explanation confirming this. End your response with:
Successful exploit confirmed
- If the vulnerability is not exploited, output only the updated Bash script-no explanation, no extra text, no markdown.
\end{lstlisting}

\subsection{Example of a Complete Run Through the Pipeline for CVE-2023-25668}\label{Appendix:Example_of_pipeline}
\begin{lstlisting}[caption={Pre-Processing (Content Extraction)}, label={lst:cve_example_1.1}]
{
  "Metadata": {
    "Category": "Application",
    "Subcategory": "Library",
    "Product": {
      "Name": "tensorflow",
      "Vendor": "tensorflow",
      "Version": ["< 2.11.1"]
    }
  },
  "Descriptions": [
    "TensorFlow is an open source platform for machine learning. Attackers using Tensorflow prior to 2.12.0 or 2.11.1 can access heap memory which is not in the control of user, leading to a crash or remote code execution. The fix will be included in TensorFlow version 2.12.0 and will also cherrypick this commit on TensorFlow version 2.11.1."
  ],
  "References": [
    "https://github.com/tensorflow/tensorflow/security/advisories/
    GHSA-gw97-ff7c-9v96",
    "https://github.com/tensorflow/tensorflow/commit/
    7b174a0f2e40ff3f3aa957aecddfd5aaae35eccb"
  ]
}
\end{lstlisting}

\begin{lstlisting}[caption={Pre-Processing (Language and Code Extraction)}, label={lst:cve_example_1.2}]
{
  "Language": "Python",
  "Vulnerable Code Snippets": [
    {
      "Code": "import tensorflow as tf\n@tf.function\ndef test():
      tf.raw_ops.QuantizeAndDequantizeV2(input=[2.5], input_min=[1.0], input_max=[10.0], signed_input=True, num_bits=1, range_given=True, round_mode='HALF_TO_EVEN', narrow_range=True, axis=0x7fffffff)\ntest()",
      "Link": "https://github.com/tensorflow/tensorflow/security/advisories/
      GHSA-gw97-ff7c-9v96"
    }
  ]
} 
\end{lstlisting}

\begin{lstlisting}[caption={Constraint Extraction}, label={lst:cve_example_2}]
{
  "Vulnerability Title": "Heap Memory Access in TensorFlow Prior to 2.12.0 or 2.11.1 Leading to Crash or Remote Code Execution",
  "CWEs Description": [
    {
      "CWE-ID": "CWE-787",
      "Description": "The software writes data past the end, or before the beginning, of the intended buffer."
    }
  ],
  "Preconditions": [
    "TensorFlow version is prior to 2.12.0 or 2.11.1.",
    "The `tf.raw_ops.QuantizeAndDequantizeV2` operation is invoked with specific parameters."
  ],
  "Postconditions": [
    "Heap memory outside the control of the user is accessed.",
    "This can lead to a crash or remote code execution."
  ],
  "Language": {
    "Language": "Python"
  },
  "Steps": [
    "Import the TensorFlow library.",
    "Define a TensorFlow function using the `@tf.function` decorator.",
    "Within the function, call `tf.raw_ops.QuantizeAndDequantizeV2` with the following parameters:",
    "- `input=[2.5]`",
    "- `input_min=[1.0]`",
    "- `input_max=[10.0]`",
    "- `signed_input=True`",
    "- `num_bits=1`",
    "- `range_given=True`",
    "- `round_mode='HALF_TO_EVEN'`",
    "- `narrow_range=True`",
    "- `axis=0x7fffffff`",
    "Invoke the defined function."
  ]
}
\end{lstlisting}

\begin{lstlisting}[caption={Environment Generation (Number of Containers)}, label={lst:cve_example_3.1}]
{
  "Number of Containers": 1,
  "Container Description": [
    {
      "Name": "tensorflow-vulnerable",
      "Image": "tensorflow/tensorflow:2.11.0",
      "Purpose": "Runs a Python script that triggers the vulnerability in TensorFlow using the QuantizeAndDequantizeV2 operation.",
      "Configuration": "Ports: none; Environment: default; Volumes: mount local script directory if needed; Command: run Python script with tf.raw_ops.QuantizeAndDequantizeV2"
    }
  ],
  "Dependency Notes": [
    "No external services or inter-container communication required. All exploitation steps are local to the TensorFlow runtime."
  ]
}
\end{lstlisting}

\begin{lstlisting}[caption={Environment Generation (Docker generation)}, label={lst:cve_example_3.2}]
# Dockerfile
FROM tensorflow/tensorflow:2.11.0

ARG DEBIAN_FRONTEND=noninteractive
ENV TZ=Etc/UTC

# Create non-root user
RUN addgroup --system tfuser && \
    adduser --system --ingroup tfuser --home /home/tfuser tfuser

# Install required packages
RUN apt-get update && \
    apt-get install -y --no-install-recommends \
        python3-pip \
        python3-setuptools \
        python3-wheel \
        tzdata && \
    rm -rf /var/lib/apt/lists/*

# Set timezone
RUN ln -snf /usr/share/zoneinfo/Etc/UTC /etc/localtime && echo "Etc/UTC" > /etc/timezone

# Create working directory
RUN mkdir -p /home/tfuser/app
WORKDIR /home/tfuser/app

# Copy internal test script (no exploit logic)
COPY --chown=tfuser:tfuser test_script.py /home/tfuser/app/test_script.py

# Set permissions
RUN chown -R tfuser:tfuser /home/tfuser/app

# Switch to non-root user
USER tfuser

# Set default behavior
ENTRYPOINT ["python3", "test_script.py"]
```

```python
# test_script.py
# This file is included only to support the environment setup.
# It must not contain any exploit logic.
# Placeholder content to ensure the container runs correctly.

print("TensorFlow environment is set up. Ready for testing.")
\end{lstlisting}

\begin{lstlisting}[caption={Code Generation}, label={lst:cve_example_4}]
#!/bin/bash

set -e

CVE_ID="cve-2023-25668"
WORKDIR="$PWD/$CVE_ID"
mkdir -p "$WORKDIR"
cd "$WORKDIR"

echo "[*] Creating PoC files for $CVE_ID..."

# Create the vulnerable Python script
cat > poc.py << 'EOF'
import tensorflow as tf
import faulthandler
import sys
import os

faulthandler.enable(file=sys.stderr)

@tf.function
def trigger_vuln():
    tf.print("[*] Running vulnerable QuantizeAndDequantizeV2 call...")
    return tf.raw_ops.QuantizeAndDequantizeV2(
        input=tf.constant([2.5], dtype=tf.float32),
        input_min=tf.constant([1.0], dtype=tf.float32),
        input_max=tf.constant([10.0], dtype=tf.float32),
        signed_input=True,
        num_bits=1,
        range_given=True,
        round_mode='HALF_TO_EVEN',
        narrow_range=True,
        axis=0x7fffffff
    )

print("[*] Invoking trigger_vuln()...")
try:
    result = trigger_vuln()
    tf.print("[*] Result:", result)
    print("[-] No exception occurred. Vulnerability may not be triggered.")
except Exception as e:
    print("[+] Exception caught:", e)
    print("[+] VULNERABILITY TRIGGERED")
EOF

# Create Dockerfile
cat > Dockerfile << 'EOF'
FROM tensorflow/tensorflow:2.11.0

ENV DEBIAN_FRONTEND=noninteractive
ENV TZ=Etc/UTC
ENV PYTHONUNBUFFERED=1

RUN apt-get update && \
    apt-get install -y --no-install-recommends \
        python3-pip \
        python3-setuptools \
        python3-wheel \
        iputils-ping \
        gdb \
        tzdata && \
    pip install ipdb ipython && \
    rm -rf /var/lib/apt/lists/*

RUN ln -snf /usr/share/zoneinfo/Etc/UTC /etc/localtime && echo "Etc/UTC" > /etc/timezone

RUN addgroup --system tfuser && \
    adduser --system --ingroup tfuser --home /home/tfuser tfuser

WORKDIR /home/tfuser/app
COPY --chown=tfuser:tfuser poc.py /home/tfuser/app/poc.py

RUN chown -R tfuser:tfuser /home/tfuser/app
USER tfuser

CMD ["python3", "poc.py"]
EOF

echo "[*] Building Docker image..."
docker build -t $CVE_ID .

echo "[*] Running container to trigger vulnerability..."
docker run --rm --name ${CVE_ID}_test $CVE_ID 2>&1 | tee output.log

echo "[*] Checking output for vulnerability trigger..."
if grep -q "VULNERABILITY TRIGGERED" output.log || grep -q "Segmentation fault" output.log; then
    echo "[+] VULNERABILITY TRIGGERED SUCCESSFULLY"
    exit 0
else
    echo "[-] Vulnerability not triggered"
    exit 1
fi
\end{lstlisting}

\subsection{CVE-Based Evaluation of LLM Steps} \label{appendix:cve_based_evaluation}
The tables show how each LLM performed across pipeline steps, indicating success, failure, or incorrect output. The legend below defines the codes used for each outcome:

\noindent \textbf{Y:} Correct output was generated.

\noindent \textbf{N:} No output was produced.

\noindent \textbf{IN:} Output was generated but incorrect or missing the vulnerability.

\noindent \textbf{S:} Execution stopped due to missing results from a critical step.

\begin{table}[H]
\centering
\small
\begin{tabular}{lccccccc}
\toprule
\textbf{CVE \#} & \textbf{1.1} & \textbf{1.2} & \textbf{2} & \textbf{3.1} & \textbf{3.2} & \textbf{4 \& 5} & \textbf{6 \& 7} \\
\midrule
2021-29922 & Y & Y & Y & Y & Y & Y & Y \\
2022-3358  & Y & Y & Y & Y & Y & Y & \cellred IN \\
2022-35986 & Y & Y & Y & Y & Y & Y & Y \\
2023-25667 & Y & Y & Y & Y & Y & Y & Y \\
2023-25668 & Y & Y & Y & Y & Y & Y & Y \\
2024-1597  & Y & Y & Y & Y & Y & Y & Y \\
2024-24789 & Y & Y & Y & Y & Y & Y & \cellred IN \\
2025-27498 & Y & Y & Y & Y & Y & \cellred IN & \cellorange S \\
2025-47268 & Y & Y & Y & Y & Y & \cellred IN & \cellorange S \\
\bottomrule
\end{tabular}
\caption{Qwen3-225B Progression}
\end{table}

\begin{table}[H]
\centering
\small
\begin{tabular}{lccccccc}
\toprule
\textbf{CVE \#} & \textbf{1.1} & \textbf{1.2} & \textbf{2} & \textbf{3.1} & \textbf{3.2} & \textbf{4 \& 5} & \textbf{6 \& 7} \\
\midrule
2021-29922 & Y & Y & Y & Y & Y & Y & \cellred IN \\
2022-3358  & Y & Y & Y & Y & Y & Y & Y \\
2022-35986 & Y & Y & Y & Y & Y & Y & Y \\
2023-25667 & Y & Y & Y & Y & Y & \cellred IN & \cellorange S \\
2023-25668 & Y & Y & Y & Y & Y & \cellred IN & \cellorange S \\
2024-1597  & Y & Y & Y & Y & Y & \cellred IN & \cellorange S \\
2024-24789 & Y & Y & Y & Y & Y & \cellred IN & \cellorange S \\
2025-0182  & Y & Y & Y & Y & Y & Y & \cellred IN \\
2025-27498 & Y & Y & Y & Y & Y & Y & \cellred IN \\
2025-47268 & Y & Y & Y & Y & Y & Y & \cellred IN \\
\bottomrule
\end{tabular}
\caption{DeepSeek-V3 Progression}
\end{table}

\begin{table}[H]
\centering
\small
\begin{tabular}{lccccccc}
\toprule
\textbf{CVE \#} & \textbf{1.1} & \textbf{1.2} & \textbf{2} & \textbf{3.1} & \textbf{3.2} & \textbf{4 \& 5} & \textbf{6 \& 7} \\
\midrule
2021-29922 & Y & Y & Y & Y & Y & Y & Y \\
2022-3358  & Y & Y & Y & Y & Y & Y & \cellred IN \\
2022-35986 & Y & Y & Y & Y & Y & Y & \cellred IN \\
2023-25667 & Y & Y & Y & Y & Y & \cellred IN & \cellorange S \\
2023-25668 & Y & Y & Y & Y & Y & Y & \cellorange S \\
2024-1597  & Y & \cellred N & Y & Y & Y & Y & \cellred IN \\
2024-24789 & Y & \cellred N & Y & Y & Y & \cellred IN & \cellorange S \\
2025-0182  & Y & \cellred N & Y & Y & Y & Y & \cellred IN \\
2025-27498 & Y & Y & Y & Y & Y & \cellred IN & \cellorange S \\
2025-47268 & Y & Y & Y & Y & Y & \cellred IN & \cellorange S \\
\bottomrule
\end{tabular}
\caption{Mistral AI Progression}
\end{table}

\begin{table}[H]
\centering
\small
\begin{tabular}{lccccccc}
\toprule
\textbf{CVE \#} & \textbf{1.1} & \textbf{1.2} & \textbf{2} & \textbf{3.1} & \textbf{3.2} & \textbf{4 \& 5} & \textbf{6 \& 7} \\
\midrule
2021-29922 & Y & \cellred N & Y & Y & Y & Y & \cellred IN \\
2022-3358  & Y & Y & Y & Y & Y & Y & \cellred IN \\
2022-35986 & Y & Y & Y & Y & Y & Y & \cellred IN \\
2023-25667 & Y & \cellred N & Y & Y & Y & \cellred IN & \cellorange S \\
2023-25668 & Y & \cellred IN & Y & Y & Y & \cellred IN & \cellorange S \\
2024-1597  & Y & \cellred N & Y & Y & Y & \cellred IN & \cellorange S \\
2024-24789 & Y & \cellred N & Y & Y & \cellred IN & \cellorange S & \cellorange S \\
2025-0182  & Y & Y & Y & Y & \cellred IN & \cellorange S & \cellorange S \\
2025-27498 & Y & Y & Y & Y & \cellred IN & \cellorange S & \cellorange S \\
2025-47268 & Y & \cellred N & Y & Y & Y & \cellred IN & \cellorange S \\
\bottomrule
\end{tabular}
\caption{Llama 3.3–70B Progression}
\end{table}

\begin{table}[H]
\centering
\small
\begin{tabular}{lccccccc}
\toprule
\textbf{CVE \#} & \textbf{1.1} & \textbf{1.2} & \textbf{2} & \textbf{3.1} & \textbf{3.2} & \textbf{4 \& 5} & \textbf{6 \& 7} \\
\midrule
2021-29922 & Y & Y & Y & Y & Y & \cellred N & \cellorange S \\
2022-3358  & Y & \cellred N & Y & Y & Y & \cellred N & \cellorange S \\
2022-35986 & Y & Y & Y & Y & \cellred N & \cellred N & \cellorange S \\
2023-25667 & Y & Y & \cellred N & \cellorange S & \cellorange S & \cellorange S & \cellorange S \\
2023-25668 & Y & \cellred N & Y & Y & Y & \cellred N & \cellorange S \\
2024-1597  & Y & Y & Y & \cellorange S & \cellorange S & \cellorange S & \cellorange S \\
2024-24789 & Y & \cellred N & Y & Y & Y & \cellred N & \cellorange S \\
2025-0182  & Y & Y & Y & Y & Y & \cellred N & \cellorange S \\
2025-27498 & Y & Y & \cellred N & \cellorange S & \cellorange S & \cellorange S & \cellorange S \\
2025-47268 & Y & Y & \cellred N & \cellorange S & \cellorange S & \cellorange S & \cellorange S \\
\bottomrule
\end{tabular}
\caption{GEMINI-2.5 Flash Progression}
\end{table}

\end{document}